\documentclass[12pt]{article}
\usepackage{psfig,epsfig}

\newcommand{\dnn}{\mbox{$\Delta N_\nu$}}
\newcommand{\beq}{\begin{equation}}
\newcommand{\eeq}{\end{equation}}
\newcommand{\zpr}{\mbox{$Z'$}}
\newcommand{\mzp}{\mbox{$M_{Z'}$}}

\newcommand{\upr}{\mbox{$U(1)'$}}
\def\mxth{\mathsurround=0pt }
\def\xversim#1#2{\lower2.pt\vbox{\baselineskip0pt \lineskip-.5pt
  \ialign{$\mxth#1\hfil##\hfil$\crcr#2\crcr\sim\crcr}}}
\def\simgr{\mathrel{\mathpalette\xversim >}}
\def\simle{\mathrel{\mathpalette\xversim <}}
\newcommand{\neff}{\mbox{$\Delta N_\nu$}}

\newcommand{\bd}{\begin{equation}}
\newcommand{\ed}{\end{equation}}

\newcommand{\bea}{\begin{eqnarray}}
\newcommand{\eea}{\end{eqnarray}}
\newcommand{\ba}{\begin{array}}
\newcommand{\ea}{\end{array}}
\newcommand{\nn}{\nonumber}

\begin{document}


\input{psfig.sty}
\begin{flushright}
\baselineskip=12pt
MADPH-02-1284 \\ UPR-1006T
\end{flushright}

\begin{center}
{\Large\bf Primordial Nucleosynthesis Constraints \\
on $Z'$ Properties}
\vglue 1.0cm
{\Large Vernon Barger$^{1}$, Paul Langacker$^{1,2}$ and Hye-Sung Lee$^{1}$}
\vglue 1cm
{$^1$ Department of Physics \\
University of Wisconsin, Madison, WI 53706 \\  
~ \\
$^2$ Department of Physics and Astronomy \\
University of Pennsylvania, Philadelphia, PA 19104 \\}
\end{center}

\vglue 1.0cm
\begin{abstract}

In models involving new TeV-scale \zpr \  gauge bosons, the new
\upr \ symmetry often prevents the generation of Majorana masses
needed for a conventional neutrino seesaw, leading to three superweakly
interacting
``right-handed'' neutrinos $\nu_R$, the Dirac partners of the ordinary
neutrinos. These can be produced prior to big bang nucleosynthesis by the
\zpr \ interactions, leading to a faster expansion rate and too
much $^4He$. We quantify the constraints on the \zpr \ properties
from nucleosynthesis for \zpr \ couplings motivated by a class of 
$E_6$ models parametrized by an angle $\theta_{E6}$.
The rate for the annihilation of three
approximately massless right-handed neutrinos into other particle
pairs through the \zpr \ channel is calculated.
The decoupling temperature, which  is higher than that of ordinary
left-handed neutrinos due to the large \zpr \  mass, is evaluated,
and the equivalent number of new doublet neutrinos
$\Delta N_\nu$ is obtained numerically as a function of the \zpr \ mass and 
couplings for a variety of assumptions concerning 
the $Z-Z'$ mixing angle and the
quark-hadron transition temperature $T_c$.
Except near the values of
 $\theta_{E6}$ for which the $Z'$ decouples
 from the right-handed neutrinos, the \zpr \ mass and mixing constraints
from nucleosynthesis are much more stringent than the existing laboratory
limits from searches for direct production or from precision electroweak
data, and are comparable to the ranges that may ultimately be probed
at proposed colliders.
For the case  $T_c = 150 $ MeV with the theoretically favored range
of
$Z-Z'$ mixings,
$\Delta N_\nu \simle 0.3$ for $M_{Z'} \simgr 4.3 $ TeV for any
value of $\theta_{E6}$. 
Larger mixing or larger $T_c$ often lead to
unacceptably large $\Delta N_\nu $ except near the $\nu_R$ decoupling
limit. 

\end{abstract}

\vspace{0.5cm}
\begin{flushleft}
\baselineskip=12pt
\today\\
\end{flushleft}
\newpage
\baselineskip=14pt

\section{Introduction}

Additional heavy \zpr \ gauge bosons~\cite{ell} are predicted in many
superstring~\cite{string} and grand unified~\cite{review} theories, and
also in models of dynamical symmetry breaking~\cite{DSB}.
If present at a scale of a TeV or so they could provide
a solution to the $\mu$ problem~\cite{muprob} and other problems
of the minimal supersymmetric standard model (MSSM)~\cite{general}.
Current limits from collider~\cite{explim,LEPmixing} and precision~\cite{indirect}
experiments are model dependent, but generally imply that $M_{Z'} > (500-800)$ GeV
and that the $Z-Z'$ mixing angle is smaller than a few $\times 10^{-3}$.
There are even hints of deviations in atomic parity violation~\cite{APV}\footnote{The
interpretation of these results is controversial. For recent discussion, see
\cite{interp}.}.
and the NuTeV experiment~\cite{NuTeV}, which could be an early indication of
a \zpr~\cite{hints}.
A \zpr \ lighter than a TeV or so should be observable at Run II
at the Tevatron. Future colliders should be able to observe
a \zpr \ with mass up to around 5
TeV and perform diagnostics on the couplings up to a few
TeV~\cite{collider}.

An electroweak or TeV-scale \zpr \ would have important implications
for theories of neutrino mass. If the right-handed neutrinos carry a non-zero
\upr \ charge, then the \upr \ symmetry forbids them from obtaining
a Majorana mass much larger than the \upr-breaking scale, and in particular
would forbid a conventional neutrino seesaw model~\cite{seesaw}.
In this case, it might still be possible to generate small Majorana
masses for the ordinary (active) neutrinos by some sort of
TeV-scale seesaw mechanism in which there are additional mass 
suppressions~\cite{TEVseesaw}. However, another possibility is that
there are no Majorana mass terms, and that the neutrinos have
Dirac masses which are small for some reason, such as higher
dimensional operators~\cite{HDO} or volume suppressions in theories with large
extra dimensions~\cite{LED}. In this case, the model would contain three additional
right-handed partners of the ordinary neutrinos, which would be almost massless.
Such light Dirac neutrinos (i.e., with mass less than an eV or so) in the
standard model or MSSM are essentially sterile, except for the tiny
effects associated with their masses and Higgs couplings, which are much
too small to produce them in significant numbers prior to
nucleosynthesis or in a supernova.  However,  the superweak interactions
of these states due to their coupling to a heavy
$Z'$ (or a heavy $W'$ in the $SU(2)_L \times SU(2)_R \times U(1)$ extension of the
standard model~\cite{LR}) might be sufficient to create
them in large numbers in the early
universe~\cite{steigman,steigman2,dolgov} or in a supernova~\cite{supernova}. 
In this paper, we consider the constraints following from big bang nucleosynthesis
on \zpr \ properties in a class of $E_6$-motivated models. 

It is well known that any new relativistic particle species that were present
when the temperature $T$ was a few MeV would increase the expansion rate,
leading to an earlier freeze-out of the neutron to proton ratio and therefore
to a higher $^4He$ abundance~\cite{yang,bbnreview}. Their contribution is usually
parametrized by the number \neff \ of additional neutrinos with full-strength
 weak interactions
that would yield the same contribution to the energy 
density.
The primordial $^4He$ abundance is still rather uncertain, but 
typical estimates of the upper limit on \neff \ are in the range\footnote{
The limit can be weakened by invoking an excess of $\nu_e$ with respect to
$\bar{\nu}_e$, which lowers the $n/p$ ratio.}
$\neff < (0.3-1)$~\cite{bbnreview,lisi}.
Of course, the $Z$-width does not allow more than 3 
light active neutrinos~\cite{PDG},
so \neff \ should be interpreted as 
an effective parameter describing degrees of freedom that do not couple with
full strength to the $Z$.

 In 1979, Steigman, Olive,
 and Schramm~\cite{steigman,steigman2} described the implications of
a superweakly interacting light particle, such as a right-handed neutrino
coupling to a heavy $Z'$. Because of their superweak interactions, such
particles decoupled earlier than ordinary neutrinos. As the temperature dropped
further, massive particles such as quarks, pions, and muons subsequently
annihilated, reheating the ordinary neutrinos and other particles in equilibrium,
but not the superweak particles. One must also take  into account
the  transition between the quark-gluon phase and the hadron phase.

A simple estimate of the decoupling temperature is obtained as
follows~\cite{steigman,steigman2}.
Ordinary  neutrinos have 
cross-sections $\sigma_W \propto G_W^2 T^2$, where $G_W$ is the
Fermi constant, and  interaction rates
\bd 
\Gamma_W(T) = n \left< \sigma_W v \right> \propto G_W^2 T^5,
\ed
where $n$ is the density of target particles.
The Hubble expansion parameter varies as $H \propto T^2/M_P$, 
where $M_P $ is the Planck scale, so
the decoupling temperature $T_d$ at which $\Gamma$ is equal to $H$ becomes
\bd
T_d \propto (G_W^2 M_P)^{-1/3}.
\ed
Putting in the coefficients,  $T_d (\nu_L) \approx 1 $ MeV
for the ordinary neutrinos\footnote{More detailed studies~\cite{bbnreview}
obtain $T_d (\nu_L) \sim 3$ MeV. We will obtain $T_d (\nu_R)$ by an explicit
calculation, so the difference is irrelevant for our purposes.}. 
Similarly, 
 a superweakly interacting particle such as a right-handed neutrino 
with a cross-section $\sigma_{SW} \propto G_{SW}^2 T^2$,  would decouple at
\bd
T_d (\nu_R)~ \sim \left( \frac{G_W}{G_{SW}} \right)^{2/3}  T_d (\nu_L).
\ed
If in the specific model, the effective superweak coupling constant $G_{SW}$ 
is proportional to $M_{SW}^{-2}$, where $M_{SW}$ is the mass of a superweak gauge boson,
the decoupling temperature can be written as
\bd
T_d(\nu_R)~ \sim \left( \frac{M_{SW}}{M_W} \right)^{4/3} T_d (\nu_L),
\label{simple}
\ed
where $M_W$ is the $W$ mass. It is then straightforward to calculate the dilution
by the subsequent quark-hadron transition and the
 annihilations of heavy particles, and the
corresponding
\neff \ from the superweak particles. 

Of course, the  estimate in
(\ref{simple}) is very rough. In particular, the detailed couplings of the \zpr \
to the $\nu_R$ and to all of the other relevant particles must be considered for
a precise estimate\footnote{Detailed calculations were carried out in~\cite{prev1}
for the $\eta$ model (see Section \ref{model}), in~\cite{prev2} for more
general $E_6$ models, and in~\cite{afdvn} for a model with generators
$T_{3R}$ and $B-L$. However, these studies considered only 
$\nu_R \overline{\nu_R} \leftrightarrow (e^+e^-, \nu_L \overline{\nu_L} )$. In the
present paper we include the interactions with all of the particles in
equilibrium at a given temperature. This leads to a lower $T_d (\nu_R)$
and more stringent limits. Constraints on extended technicolor models
were considered in~\cite{ETC}.}
In this paper, we do this for a class of $Z'$ models with couplings
motivated by $E_6$ grand unification~\cite{lw}. (The full structure of $E_6$
is not required.)
We define the $U(1)'$ model in Section \ref{model}. 
The implications of superweakly coupled particles for nucleosynthesis
and the uncertainties from the quark-hadron transition temperature $T_c$
are summarized  in Section \ref{nucleosynthesis}.
Section \ref{calculation} deals with the 
calculation of the decoupling temperature.
We present our results and numerical analysis for $T_d$ and \neff \
for three right-handed neutrinos as a function
of the \zpr \ mass and  couplings for various assumptions concerning the$Z-Z'$ mixing
and $T_c$ in Section
\ref{result}.  The discussion and conclusion follows in Section \ref{discussion}.
 
\section{$Z'$ in $E_6$-motivated models}
\label{model}
A general model with an extra \zpr \ is characterized by the \zpr \
mass; the $Z-Z'$ mixing angle; the \upr \ gauge coupling; the
\upr \ chiral charges for all of the fermions and scalars,
which in general may be family non-universal, leading to flavor changing
neutral currents~\cite{famnon};
and an additional parameter associated with mixing
between the $Z$ and $Z'$ kinetic terms~\cite{kinetic}.
Furthermore, most concrete \zpr \ models involve additional
particles with exotic standard model quantum numbers, which are
required to prevent anomalies. It is difficult to work with the
most general case, so many studies make use of the \upr \ charges
and exotic particle content associated with the $E_6$ model,
as an example of a consistent anomaly-free construction\footnote{
The full structure of $E_6$ grand unification is not required, and in
fact the $E_6$ Yukawa coupling relations must not  be respected
in order to prevent rapid proton decay~\cite{lw}.}. Explicit
string constructions~\cite{stringmodels} often lead to other
patterns of couplings and exotics, but these are very model
dependent.

$E_6$ actually yields two additional \upr \ factors when broken
to the standard model (or to $SU(5)$), i.e.,
\bd
E_6 \to SO(10) \times U(1)_\psi \to SU(5) \times U(1)_\chi \times U(1)_\psi.
\ed
It is usually assumed that only one linear combination survives to
low energies, parametrized by a mixing angle $\theta_{E6}$.
The resultant $U(1)'$ charge is then\footnote{We ignore the possibility of
kinetic mixing~\cite{kinetic}.}
\bd
Q = Q_\chi \cos \theta_{E6} + Q_\psi \sin \theta_{E6}. \label{Qdef}
\ed
A special case that is often considered is $U(1)_\eta$, which
corresponds to $\theta_{E6} = 2 \pi - \tan^{-1} \sqrt{\frac{5}{3}}=1.71 \pi$.
We list the charges of $U(1)_\chi$ and $U(1)_\psi$ that we 
need in Table \ref{table1}.
The quantum numbers of the associated exotic particles are given in~\cite{lw}.
It is conventional to choose $\theta_{E6}$ to be in the range $(0,\pi)$,
since the charges merely change sign for  $\theta_{E6} \rightarrow  \theta_{E6} + \pi$.
With this convention one must allow both positive and negative values for the
$Z-Z'$ mixing angle $\delta$. In this paper, we find it convenient
to choose a different convention in which $\theta_{E6}$ varies from $0$ to $2 \pi$, but
for which $\delta \le 0$.  That is, the range $0-\pi$ corresponds to the
$E_6$ models with negative mixing, while $\pi-2 \pi$ corresponds to
positive mixing.
The $\nu_R$ charge is nonzero, precluding
an ordinary seesaw, except for $\theta_{E6} \sim 0.42 \pi$ and $1.42 \pi$.
We will always assume that the neutrinos are Dirac and that the three
right-handed neutrinos are therefore very light. (In fact, the non-zero Dirac
masses play no role in the analysis.)
There could be additional sterile states,
such as the $SO(10)$-singlet states occurring in the 27-plet of $E_6$. If these
involve nearly-massless fermions they could also contribute to the expansion rate prior
to nucleosynthesis. We  assume that these additional neutralinos
acquire electroweak scale masses from the gauge symmetry breaking~\cite{ell}.

\renewcommand{\arraystretch}{1.4}
\begin{table}[t]
\caption{The (family-universal) charges of the $U(1)_\chi$ and the $U(1)_\psi$.}
\label{table1}
\vspace{0.4cm}
\begin{center}
\begin{tabular}{|c|c|c|}
\hline
Fields & $Q_\chi$ & $Q_\psi$ \\
\hline
$u_L$ & $-1/2 \sqrt{10}$ & $1/2 \sqrt{6}$ \\
\hline
$u_R$ & $1/2 \sqrt{10}$ & $-1/2 \sqrt{6}$ \\
\hline
$d_L$ & $-1/2 \sqrt{10}$ & $1/2 \sqrt{6}$ \\
\hline
$d_R$ & $-3/2 \sqrt{10}$ & $-1/2 \sqrt{6}$ \\
\hline
$e_L$ & $3/2 \sqrt{10}$ & $1/2 \sqrt{6}$ \\
\hline
$e_R$ & $1/2 \sqrt{10}$ & $-1/2 \sqrt{6}$ \\
\hline
$\nu_L$ & $3/2 \sqrt{10}$ & $1/2 \sqrt{6}$ \\
\hline
$\nu_R$ & $5/2 \sqrt{10}$ & $-1/2 \sqrt{6}$ \\
\hline
\end{tabular}
\end{center}
\end{table}

Let $Z$ and \zpr \ represent the Standard Model and \upr \ gauge bosons, respectively,
and  $Z_{1,2}$  the mass eigenstate bosons, related by
\bd
\left( \begin{array}{c} Z_1 \\ Z_2 \end{array} \right) =
\left( \begin{array}{cc} \cos \delta  &
 -\sin \delta \\ \sin \delta & \cos \delta \end{array}  \right) 
\left( \begin{array}{c} Z \\ Z' \end{array}  \right),
\ed
where $\delta$ is the $Z-Z'$ mixing angle. As stated in the Introduction, the limits
on $M_{Z_2} \sim M_{Z'}$ depend on $\theta_{E6}$ and also on the masses of any exotics and
superpartners to which the $Z'$ couples,
but are typically in the range $\mzp > (500 - 800)$ GeV.
The limits on $\delta$ are correlated with those for \mzp \ and are asymmetric under
$\delta \rightarrow - \delta$. However, for $\mzp \sim 1$ TeV
the constraints are less sensitive to $\theta_{E6}$ and are approximately  symmetric,
with $|\delta|<0.002$ giving a reasonable approximation for all $\theta_{E6}$.
For larger \mzp \ there are two theoretical constraints on the mixing,
corresponding to equations (6) and (5) of~\cite{luo}. The first 
is a theoretical relation between the
mass and mixing,
\beq
\delta = C \frac{g_{Z}'}{g_Z}  \frac{M_{Z_1}^2}{M_{Z_2}^2}, \label{mixval} \eeq
where 
$g_Z \equiv \sqrt{g_1^2 + g_2^2}$
and $g_Z'$ is the  \upr \ gauge coupling constant. The value of $g_Z'$
depends on the embedding and breaking of the underlying theory. We will choose
$g_Z' = \sqrt\frac{5}{3} g_Z \sin \theta_W$, which corresponds to a unification of
$g_Z'$ with the other gauge couplings for the exotic particle quantum numbers of
supersymmetric $E_6$.
In (\ref{mixval}) $C$ depends on the charges of the scalar fields which
lead to the mixing (see Table III of~\cite{luo}). However, for the
typical cases in which the mixing is induced by scalars 
in an $E_6$ 27 or $\overline{27}$-plet,
 it is a reasonable approximation
to take $-1 < C < 1$ for all $\theta_{E6}$. (One can have a slightly
more restrictive range for some $\theta_{E6}$.)
The assumption $|C| < 1$ corresponds to
$|\delta | < 0.0051/M_{Z_2}^2$,
where $M_{Z_2}$ is  in TeV.
The second theoretical constraint
is the
requirement that the mixing should not change the mass of the lighter $Z$  more
than is allowed by the data. It is equivalent to
\beq
| \delta | \sim \sqrt{\rho_0 -1 }\ \  \frac { M_{Z_1}}{M_{Z_2}},  \eeq
where
$M_{Z_1} = M_Z$, and  the $\rho_0$ parameter, defined precisely
in~\cite{erler2}, should be exactly 1 in the standard model. The precision data
imply $\rho_0 < 1.001$. Hence,
$|\delta | < 0.0029/M_{Z_2}$,
where $M_{Z_2}$ is again in TeV. 
We will consider the following cases: 
\bea
{\rm (A0)}\ \ \delta \ &=& 0 \ \  {\rm (no\ mixing)} \nn \\
{\rm (A1)}\
|\delta | &<& 0.0051/M_{Z_2}^2 \ \ 
      {\rm (theoretical \ mass-mixing \ relation)} \nn
\\ 
{\rm (A2)}\ |\delta | &<& 0.0029/M_{Z_2} \ \  {\rm (\rho_0 \ constraint)} \nn \\
{\rm (A3)}\ |\delta | &=&  0.002 \ \ {\rm (maximal\ mixing\ allowed\ for\ M_{Z_2} \sim 1
\ TeV)}.
\label{cases}
\eea
A1 is more stringent than A2 and A3 in the large mass range, so we will mainly
focus on A0 and A1.

The lagrangian for the massive neutral current coupling to fermion $f$ is~\cite{luo}
\bea
{- \mathcal L}_{\it int} &=&  g_Z  Q_Z(f_L){\bar f_L}  \gamma^\mu 
f_L Z_\mu + g_Z Q_Z(f_R){\bar f_R}  \gamma^\mu f_R Z_\mu \nn \\
& & + g_Z' Q(f_L){\bar f_L} 
\gamma^\mu f_L Z'_\mu + g_Z' Q(f_R){\bar f_R}  \gamma^\mu f_R Z'_\mu 
\eea
where 
\bea
Q_Z(f_L) \equiv T^3_f - q_f \sin^2 \theta_W, \nn \\
Q_Z(f_R)\equiv - q_f \sin^2 \theta_W, 
\eea
and $Q(f_{L,R})$ is given by (\ref{Qdef}).
The annihilation cross-section through $Z'$ has both (light) $Z_1$ and (heavy)
$Z_2$ contributions unless $\delta = 0$ and is calculated in Section \ref{calculation}.

\section{Nucleosynthesis}
\label{nucleosynthesis}
As described in the Introduction, the
observed $^4He$ abundance constrains the energy density at the time
of Big Bang Nucleosynthesis \cite{yang}, with most recent
estimates~\cite{bbnreview,lisi} of the number of equivalent new
active neutrino types in the range $\neff < (0.3-1)$. 

The contribution of new relativistic species can be written
\bd
\Delta N_\nu = \frac{8}{7} \sum_B \frac{g_B}{2} 
\left( \frac{T_B}{T_{BBN}} \right)^4 + \sum_F \frac{g_F}{2} \left( \frac{T_F}{T_{BBN}} \right)^4,
\ed
where $g_B$ and $g_F$ are degrees of freedom of new bosons (B) and new fermions (F), respectively,
$T_{B,F}$ are their effective temperatures, and $T_{BBN} \sim 1$  MeV is the
temperature at the time of the freeze-out of the neutron to proton ratio.
In particular, the contribution of three types of right-handed neutrinos is
\beq \label{NfromT}
\Delta N_\nu  = 3 \cdot 1 \cdot \left( \frac{T_{\nu_R}}{T_{BBN}} \right)^4
= 3 \left( \frac{g(T_{BBN})}{g(T_d(\nu_R))} \right)^{4/3}
\eeq
where $T_d(\nu_R)$ is the decoupling temperature of the right-handed neutrinos.
$g(T)$ is the effective number of degrees of freedom at temperature $T$.
Neglecting finite mass corrections, it is given by $g_B(T) + \frac{7}{8} g_F(T)$,
where $g_{B,F}(T)$ are the number of bosonic and fermionic relativistic degrees
of freedom in equilibrium at temperature $T$~\cite{steigman,steigman2}.
In particular, 
$g(T_{BBN}) = 43/4$ from the three active neutrinos, $e^{\pm}$, and $\gamma$, and
$g(T)$ increases  (in this approximation)
as a series of step functions at higher temperature as more
particles are in equilibrium. 
The second equality in (\ref{NfromT}) comes 
from entropy conservation \cite{steigman}
in the heavy particle decouplings and quark-hadron transition  subsequent to
the $\nu_R$ decoupling. Therefore, 
the $\nu_R$ are  not included in our definition of $g(T)$.
(They will be included in the expansion rate formula prior to decoupling.)

In calculating $g(T)$ 
one must also take into account the QCD phase transition
at temperature $T_c$. Above $T_c$ the $u$ and $d$ (and possibly $s$)
quarks and the gluons were the relevant
hadronic degrees of freedom,
while below $T_c$ they are replaced by pions~\cite{steigman,steigman2}.
The value of $T_c$ is poorly known, but is usually estimated to be in the range
$(150-400)$ MeV~\cite{olive}. This range is estimated in quark and hadron
potential models as the temperature above which hadrons start to overlap
(lower end) or as the temperature below which the quark gas in no longer
ideal (upper end). A related uncertainty is whether to use current or
constituent quark masses. At very high temperatures the quarks
can be considered as asymptotically free and current masses are appropriate, while
around $T_c$
constituent effects become important\footnote{One can alternatively
argue that the current masses are appropriate above a temperature $T_{chiral}$,
above which chiral symmetry is restored, 
and constitutent masses below  $T_{chiral}$.
One would expect  $T_c$ and  $T_{chiral}$ to be comparable, but their precise 
relation is uncertain.}.
The range of estimates
for $T_c$ is essentially unchanged if one simply fixes the quark masses
at either value~\cite{olive}.   

\begin{figure}
\begin{center}
\resizebox{!}{3.0in}{\includegraphics{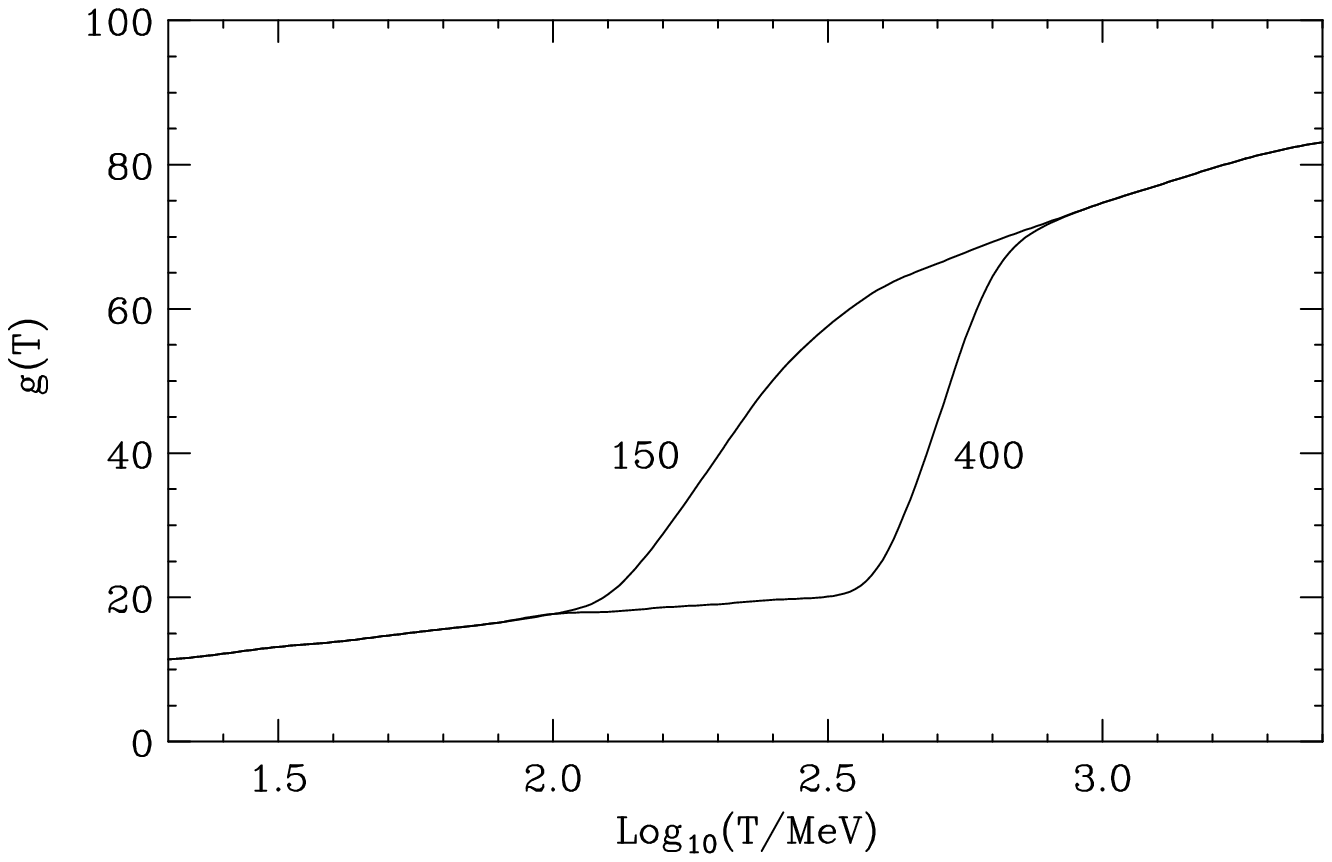}}
\end{center}
\caption{The effective number of degrees of freedom as a function of temperature for the
quark-hadron  transition temperature $T_c=
150$ MeV and $400$ MeV, from~\cite{srednicki}.
$g(T)$ does not include  contributions from
 the three right-handed neutrinos, which are added separately 
in the expansion rate formula.}
\label{dof}
\end{figure}

Figure \ref{dof} shows the explicit values of 
$g(T)$ from the more detailed analysis of
 Ref.~\cite{srednicki}, which includes finite mass and other corrections,
and uses the two values
$T_c= 150$ MeV and 400 MeV. We will also use these 
values for our numerical analysis.
The sharp increase in $g(T)$ above $T_c$ (because of the large number
of quark and gluon degrees of freedom) is extremely important for relaxing
the constraints on the \zpr \ mass.

The QCD phase transition does not occur instantaneously or at one temperature
but rather smoothly (meaning both quarks and hadrons exist at the
same temperature) for a period of time around 
$T_c$, as illustrated by the smooth curves in Figure \ref{dof}.
Risking a small inconsistency,
we approximate our calculation of the interaction rate 
by simply switching from quarks
to hadrons for temperatures below $T_c$. We will take
the values $T_c$ = 150 and 400 MeV to illustrate the range of hadronic
uncertainties.
Above $T_c$, the interaction rate
depends in principle on the quark masses, especially for low $T_c$.
However, we have found in practice that the results are almost identical
for constituent and current masses, so we will mainly display them for
the constituent case (both will be shown for the $\eta$ model).

The  calculation of the right-handed neutrino decoupling temperature, 
$T_d(\nu_R)$ in terms of the $Z'$ parameters is  discussed  in the next section.

\section{The expansion and interaction rates}
\label{calculation}
A particle is decoupled from the background when its interaction rate drops below
the expansion rate of the universe.
In this section, we present the the cosmological expansion rate
$H(T)$ along with the explicit form of the interaction rate $\Gamma(T)$ 
for $\overline{\nu_R} \nu_R$ annihilating into all open
channels\footnote{As long as equilibrium is maintained, the $\nu_R$ annihilation and
production rates are the same. It is more convenient to estimate
the annihilation rate of $\overline{\nu_R} \nu_R$ into massive particles, because
the final state mass effects are easily incorporated in the cross section
formulae, whereas for the production rate one must explicitly consider
the suppressed number density for the massive particles.},
and estimate the decoupling temperature $T_d$ of a
right-handed neutrino by $\Gamma(T_d) \sim H(T_d)$. 

The Hubble expansion  parameter is given by
\bd
H(T) = \sqrt{\frac{8 \pi G_N \rho(T)}{3}}
= \sqrt{\frac{4 \pi^3 G_N g'(T)}{45}} T^2
\ed
where $G_N = M_P^{-2}$ is the Newton constant and $\rho(T)$ is
the energy density.
We define
$g'(T) = g(T) + \frac{21}{4}$, where the $21/4$
reflects the 3 massless right-handed neutrinos.

The cross-section $\sigma_i (s) \equiv \sigma(\overline{\nu_R} \nu_R \to 
\overline{ f_i} f_i)$
for a massless right-handed neutrino 
pair to annihilate into a fermion pair through the $Z'$-channel is

\bea \label{sigma}
\sigma_i (s)  = N_C^i \frac{s \beta_i}{16 \pi}
\left\{ \left( 1 + \frac{\beta_i^2}{3} \right) \left(
(G_{RL}^i)^2 + (G_{RR}^i)^2 \right) +
2 \left( 1 - \beta_i^2 \right) G_{RL}^i G_{RR}^i \right\}
\eea
where (for $s \ll M_{Z_1}^2, M_{Z_2}^2$)
\bea
G_{RX}^i &=& g_Z'^2 Q(\nu_R) Q(f_{iX}) 
\left( \frac{\sin^2 \delta}{M_{Z_1}^2} + \frac{\cos^2
\delta}{M_{Z_2}^2} \right) \nn \\
&-& g_Z' g_Z Q(\nu_R) Q_Z(f_{iX}) \left( \frac{\sin \delta \cos \delta}{M_{Z_1}^2} -
\frac{\sin \delta \cos \delta}{M_{Z_2}^2} \right),  
\label{grx}
\eea
where $X = L$ or $R$, \ \
 $\beta_i \equiv \sqrt{1 - 4 m_{f_i}^2 / s}$ is the relativistic velocity for the
 final particles,
and $N_C^i$ is the color factor of particle $f_i$.

In the limit of no-mixing $( \delta = 0 )$ and massless final particles $( \beta_i =
1 )$, the cross-section simplifies to

\beq
\sigma_i (s) 
\to N_C^i \frac{s}{12 \pi} \left( \frac{g_Z'^2}{M_{Z'}^2}
\right)^2 Q(\nu_R)^2 \left( Q(f_{iL})^2
+ Q(f_{iR})^2 \right),
\label{nomixlimit}
\eeq
consistent with  the earlier estimate $\sigma_{SW} \propto G_{SW}^2
T^2$ with $G_{SW} \propto \frac{g_Z'^2}{M_{Z'}^2}$ and $T \propto \sqrt{s}$.

For temperatures less than the  quark-hadron transition temperature
$T_c = 150 - 400$ MeV, we replace the quark degrees of freedom  with hadrons.
The only relevant annihilation channels are into charged pions.
We approximate the cross-section of $\overline{\nu_R} \nu_R$
annihilating into $\pi^+ \pi^-$ by using the
$\rho$ dominance model \cite{sakurai}.
\bea \label{sigmapi}
\sigma_\pi (s) \equiv \sigma(\overline{\nu_R} \nu_R \to 
\pi^+ \pi^-) = \frac{s \beta_\pi^3}{96 \pi} | F_\pi (s) |^2
\left( G_{RL}^u + G_{RL}^{\bar d} + G_{RR}^u + G_{RR}^{\bar d} \right)^2
\eea
which is basically obtained by using
$Q(f_{iL}) = Q(u_L) + Q(\bar d_L)$
and $Q_Z(f_{iL}) = Q_Z(u_L) + Q_Z(\bar d_L)$ for $G_{RL}^i$ and likewise for $G_{RR}^i$.
The pion form factor\footnote{More complicated form factors
are known to fit the experimental data better~\cite{sakurai2}, but
(\ref{formfactor}) is adequate for our purposes.} is
\beq
F_\pi (s) = \frac{m_\rho^2}{s - m^2_\rho + i m_\rho \Gamma_\rho}, \label{formfactor} \eeq
with
$m_\rho = 771$ MeV and  $\Gamma_\rho = 149$ MeV.

\renewcommand{\arraystretch}{1.4}
\begin{table}[t]
\caption{The masses (in MeV) used for  the numerical analysis.}
\label{table2}
\vspace{0.4cm}
\begin{center}
\begin{tabular}{|c|c||c|c|}
\hline
Quarks & Current (Constituent) masses & Others & Masses\\
\hline
$u$ & $4.2\ (340)$ & $\nu$ & $0$ \\
\hline
$d$ & $7.5\ (340)$ & $e$ & $0.511$ \\
\hline
$s$ & $150\ (540)$ & $\mu$ & $105$ \\
\hline
$c$ & $1150\ (1500)$ & $\tau$ & $1800$ \\
\hline
$b$ & $4200\ (4500)$ & $\pi$ & $137$ \\
\hline
\end{tabular}
\end{center}
\end{table}

The interaction rate per $\nu_R$ is  
\bd
\Gamma(T) = \sum_i \Gamma_i(T) = \sum_i \frac{n_{\nu_R}}{g_{\nu_R}} 
\left< \sigma v (\overline{\nu_R} \nu_R \to 
\overline{ f_i} f_i,~ \pi^+ \pi^-) \right>,
\ed
where $n_{\nu_R}$
is the number density of a single flavor of massless right-handed
 neutrinos plus antineutrinos,
 $g_{\nu_R}=2$ is the number of degrees of freedom,
 and $\left< \sigma v \right>$ is the
thermal average of the cross-section times velocity.

We use the same masses (Table \ref{table2}) used in the
calculation~\cite{steigman2,srednicki}
of $g(T)$ in Figure~\ref{dof}, except for
the value $m_b = 4200$ MeV of the $b$ quark current  mass~\cite{PDG}. 
We include the contributions of all
particles up to  the $b$ quarks.
The contributions from the top quark and 
heavy particles from new physics, such as
squarks, sleptons, and exotics would only be relevant 
when the decoupling temperature is close
to the electroweak scale or higher. This only occurs when $\theta_{E6}$ 
is extremely close to
the values for which the $\nu_R$ decouples from the \zpr.

For a massless right-handed neutrino pair colliding with 4-momenta 
$p^\mu \equiv (p, \bf p)$
and $k^\mu \equiv (k, \bf k)$ with  relative 
angle $\theta$, the interaction rate per neutrino is~\cite{Kolb}
\bea
\Gamma_i (T) &=& \frac{g_{\nu_R}}{n_{\nu_R} (T)} 
\int \frac{d^3 \bf p}{(2 \pi)^3} \frac{d^3 \bf k}{(2 \pi)^3} 
f_\nu(p) f_\nu(k) \sigma_i(s) v_M \nn
\\ &=& \frac{ g_{\nu_R}}{8 \pi^4 n_{\nu_R}} 
\int_0^\infty p^2 dp \int_0^\infty k^2 dk \int_{-1}^1 d \cos
\theta ~\frac{\left(1 - \cos \theta\right)}{\left(e^{k/T} + 
1\right) \left(e^{p/T} + 1\right)}
\sigma_i (s),
\eea
 where  
$f_\nu(k) =  (e^{k/T} + 1)^{-1}$
is the Fermi-Dirac distribution with
 \bd
n_{\nu_R} (T) = g_{\nu_R} \int \frac{d^3 \bf k}{(2 \pi)^3} 
f_\nu(k) = 2 \cdot  \frac{3}{4 \pi^2} \zeta(3) T^3,
\ed
$v_M = p \cdot k/pk = 1 - \cos\theta$
is the M\o ller velocity, and
$s = 2 p k (1 - \cos\theta)$
is the square of the center-of-mass energy.

A root-finding method was used to calculate the decoupling temperature, 
for which $H =
\Gamma$. A several percent  error  was allowed in 
the numerical result to calculate the roots
efficiently. 
Finite temperature effects, such as changes in the phase space 
due to interactions with the thermal bath, can increase the ordinary neutrino 
decoupling temperature by
several percent~\cite{fornengo}. Analogous effects for the
$\nu_R$ are too small to significantly affect our
results.

\section{Numerical results}
\label{result}
\begin{figure}
\begin{center}
\resizebox{!}{1.87in}{\includegraphics{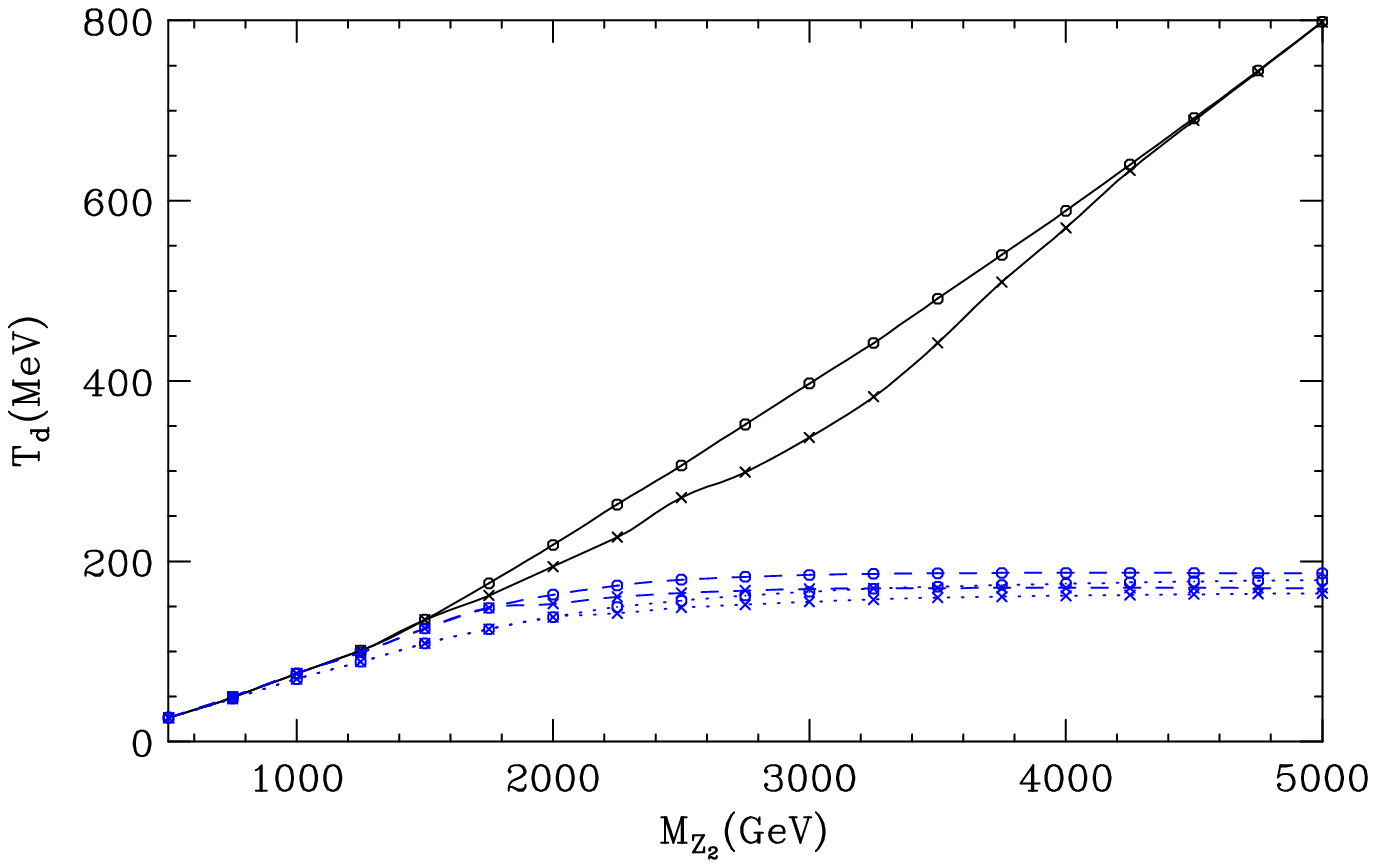}}
\resizebox{!}{1.87in}{\includegraphics{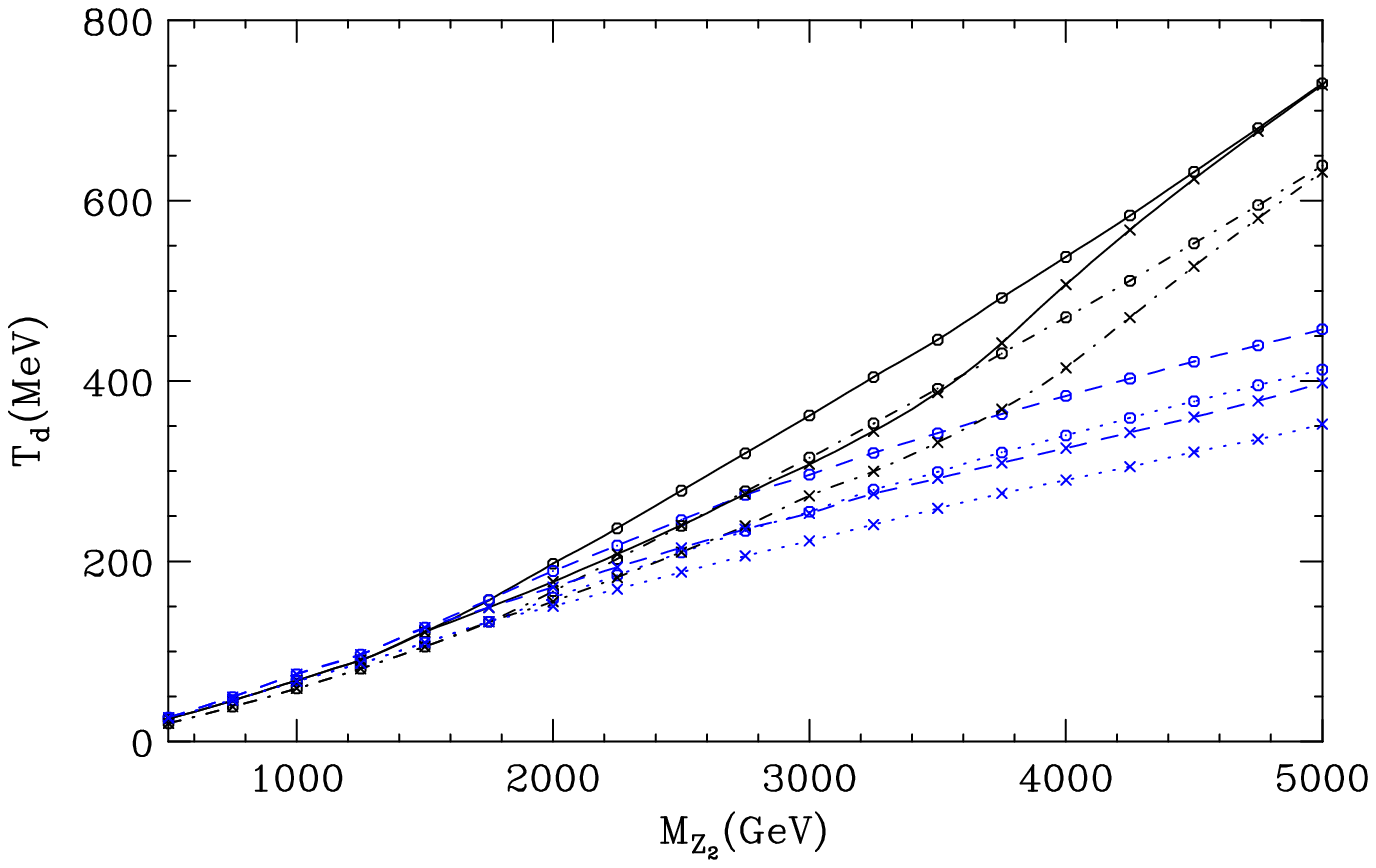}}
\resizebox{!}{1.87in}{\includegraphics{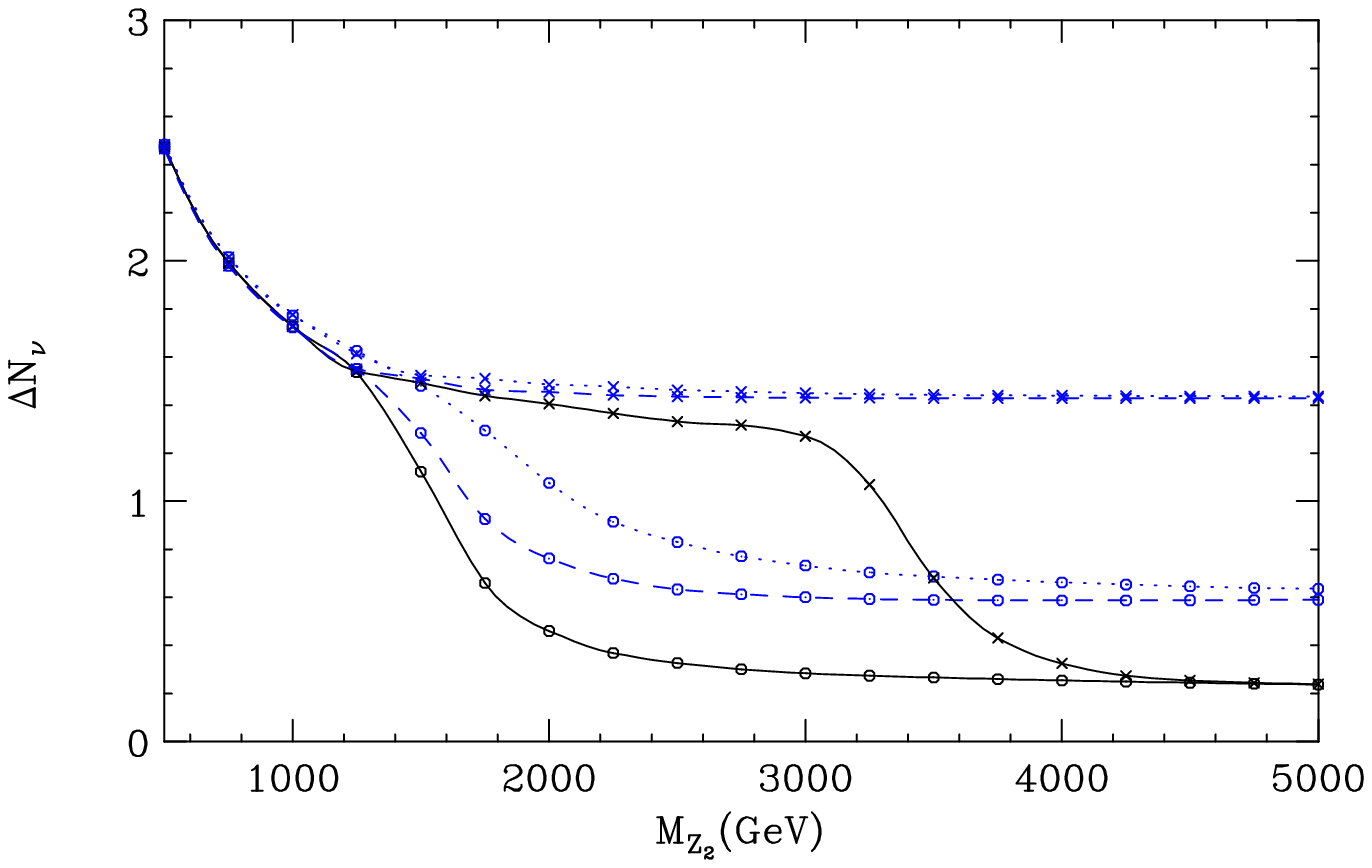}}
\resizebox{!}{1.87in}{\includegraphics{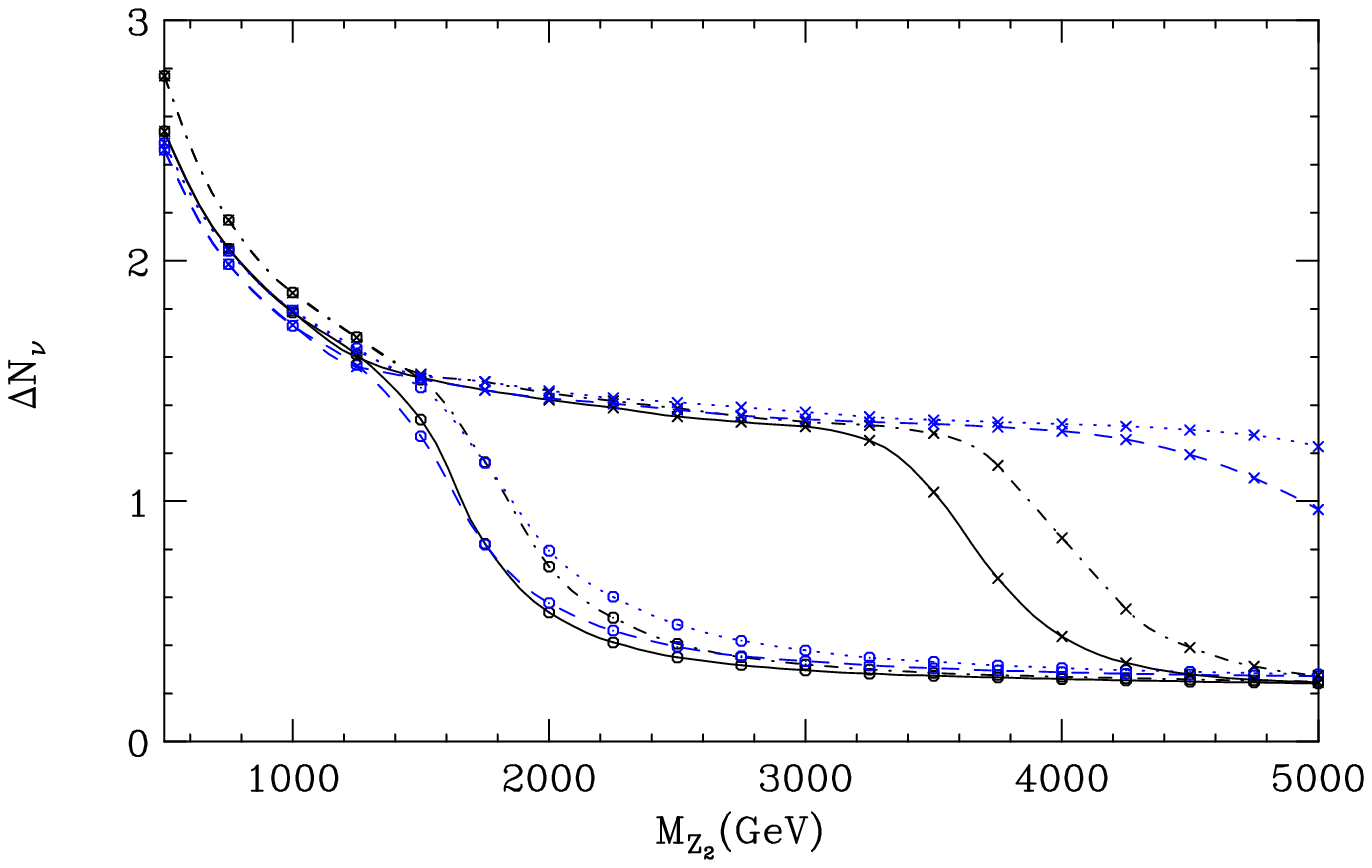}}
\end{center}
\caption{The decoupling temperature $T_d$ (top) and the equivalent number of extra
neutrinos $\Delta N_\nu$ (bottom) for the $\eta$ model 
as a function of the $Z_2$ mass $M_{Z_2}$  for constituent quark masses, for a
 quark-hadron transition temperature $T_c=$  150 MeV (circles) and
 400 MeV (crosses). The left two figures are for the cases A0 and A3 defined in (\ref{cases}),
i.e., 
 the solid, dashed and dotted
lines represent  zero-mixing ($\delta = 0$),  and positive and negative maximal-mixing  
($\delta = \pm
0.002$), respectively. The $T_c$ = 150
MeV case has higher $T_d$ and lower $\Delta N_\nu$ for the same $M_{Z_2}$ than $T_c$ = 400 MeV.
The right figures are for the intermediate mixing assumptions A1 and A2.
The solid and dash-dot curves are for the  mass-mixing relations
 $\delta= \pm 0.0051/M_{Z_2}^2$, while the dashed
and dotted curves are  for the
$\rho_0$ constraints
$\delta =\pm 0.0029/M_{Z_2}$.}
\label{decoupling1}
\end{figure}

\begin{figure}
\begin{center}
\resizebox{!}{1.87in}{\includegraphics{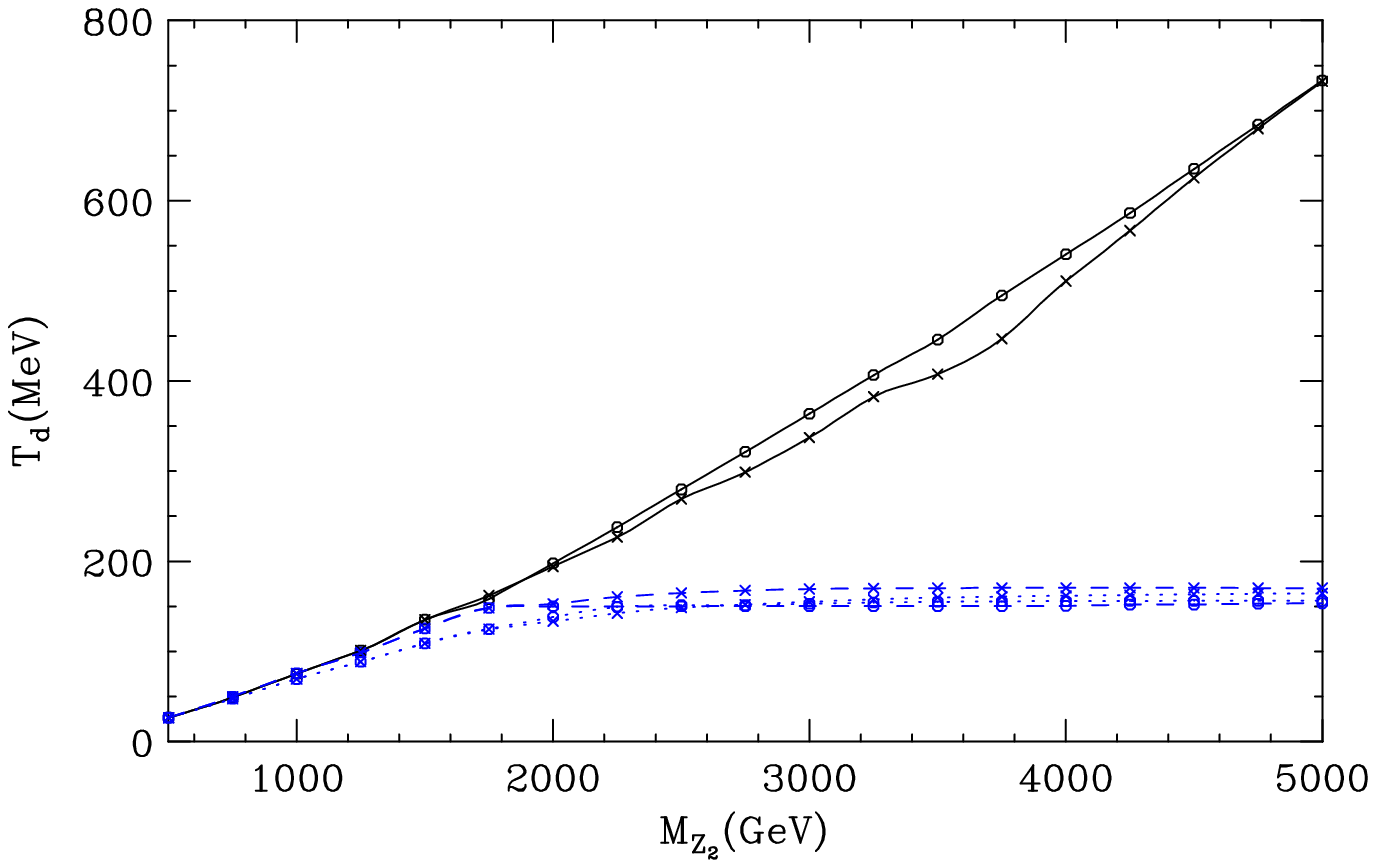}}
\resizebox{!}{1.87in}{\includegraphics{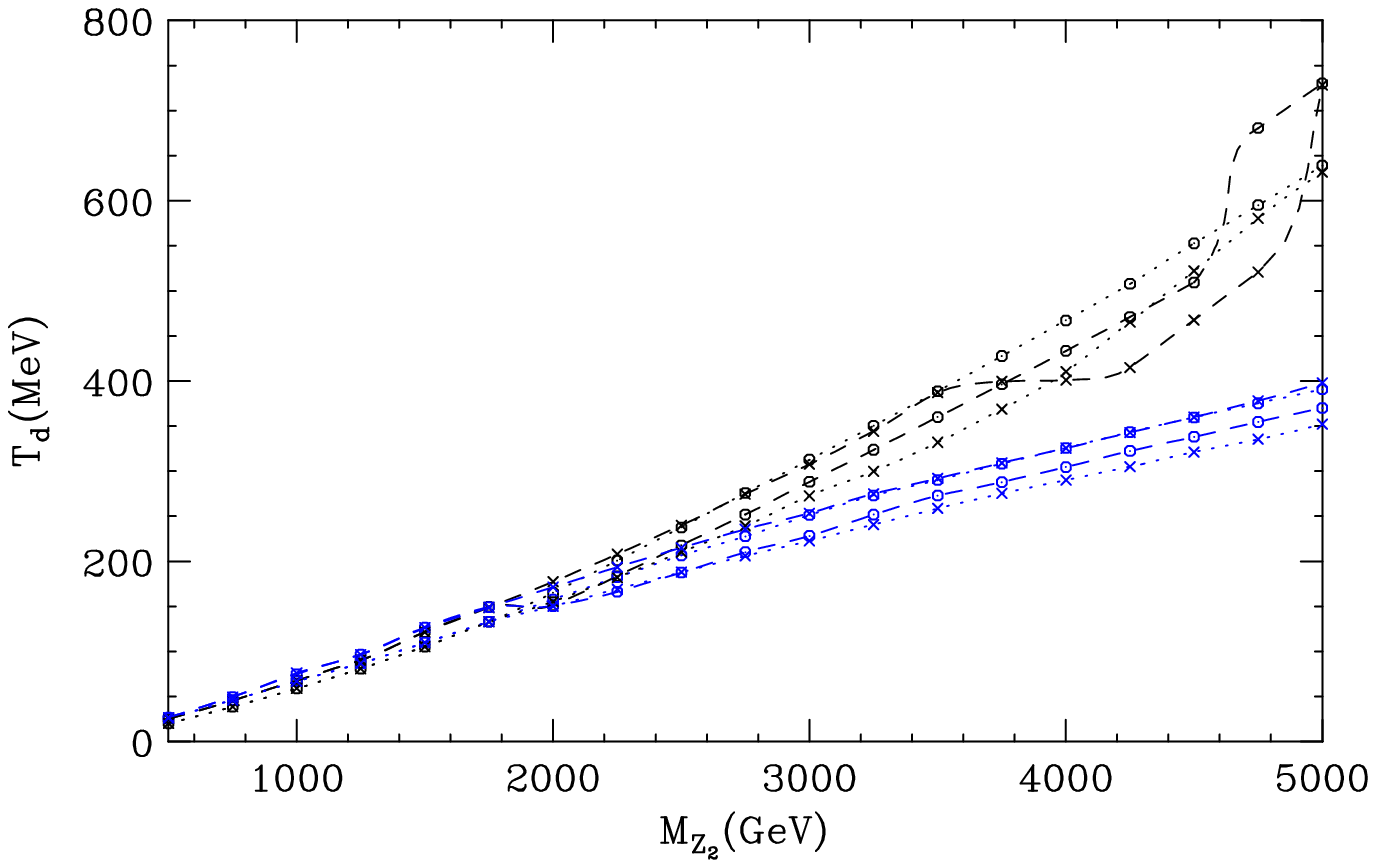}}
\resizebox{!}{1.87in}{\includegraphics{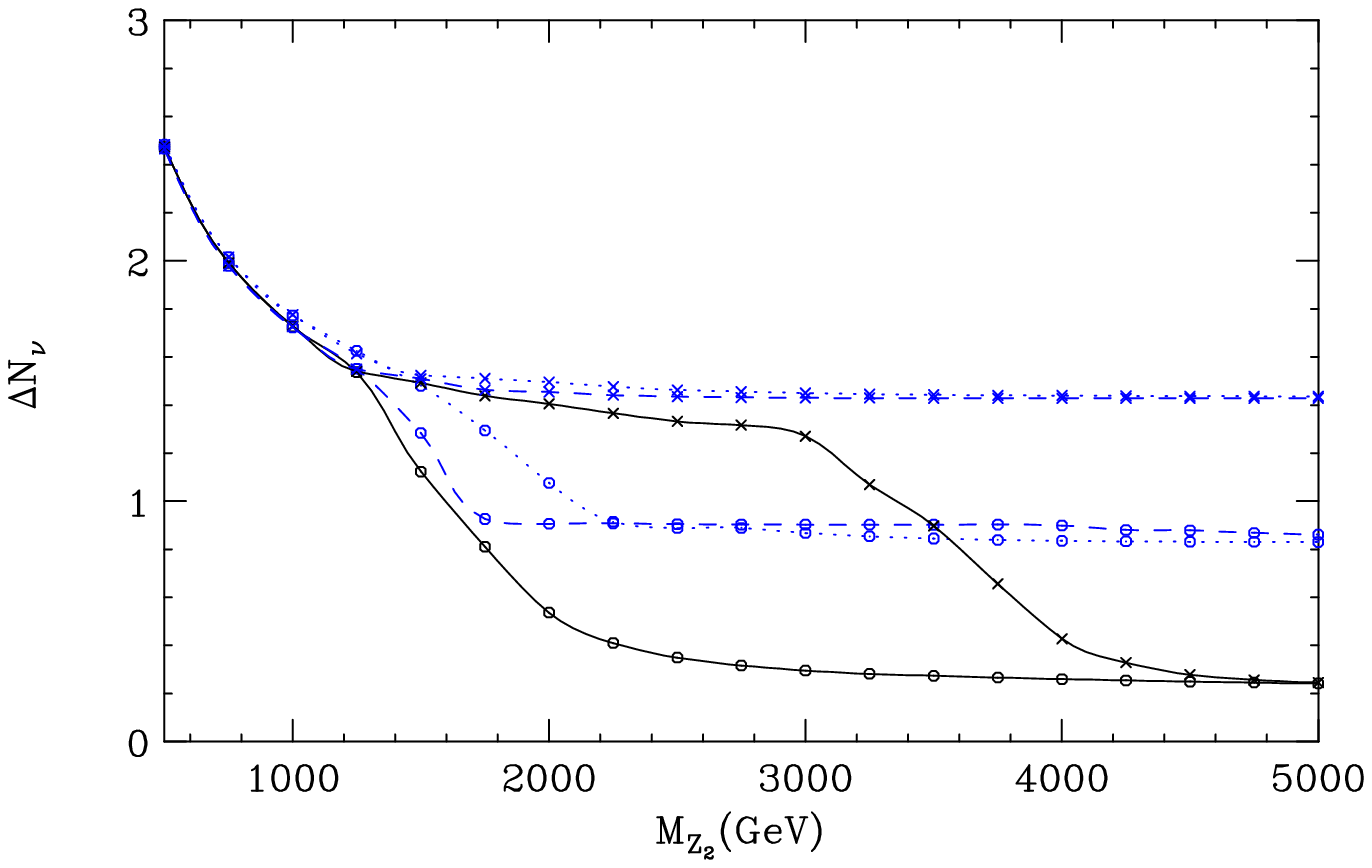}}
\resizebox{!}{1.87in}{\includegraphics{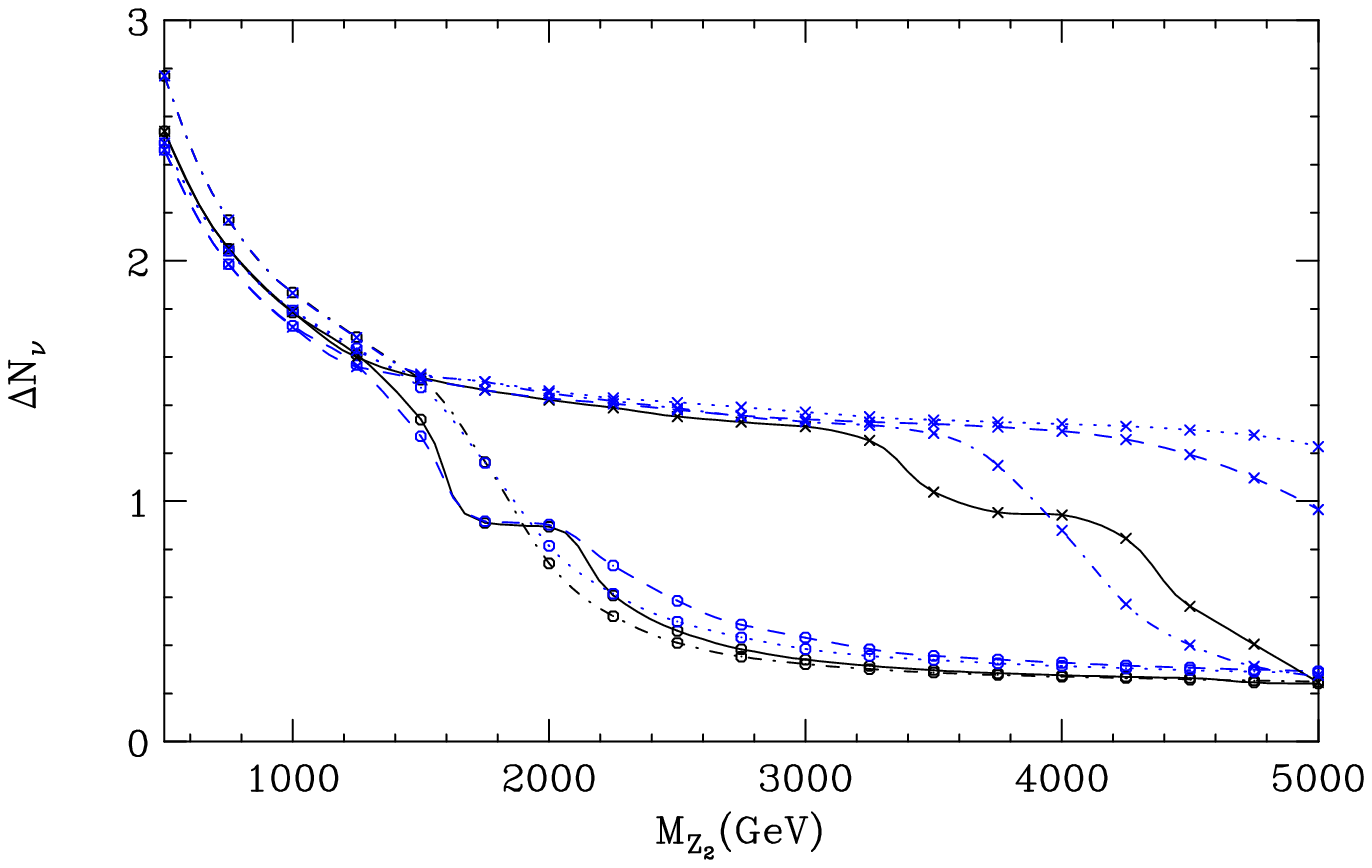}}
\end{center}
\caption{
Same  as Figure \ref{decoupling1} except that  current quark masses are used. 
The upper graphs share most features with the constituent mass case except  that $T_d$ can
be slightly lower  when $T_d > T_c$. The only significant
change in $\Delta N_\nu$ is for the $T_c$ = 150 MeV maximal mixing case (see text).}
\label{decoupling1a}
\end{figure}

\begin{figure}
\begin{center}
\resizebox{!}{2.38in}{\includegraphics{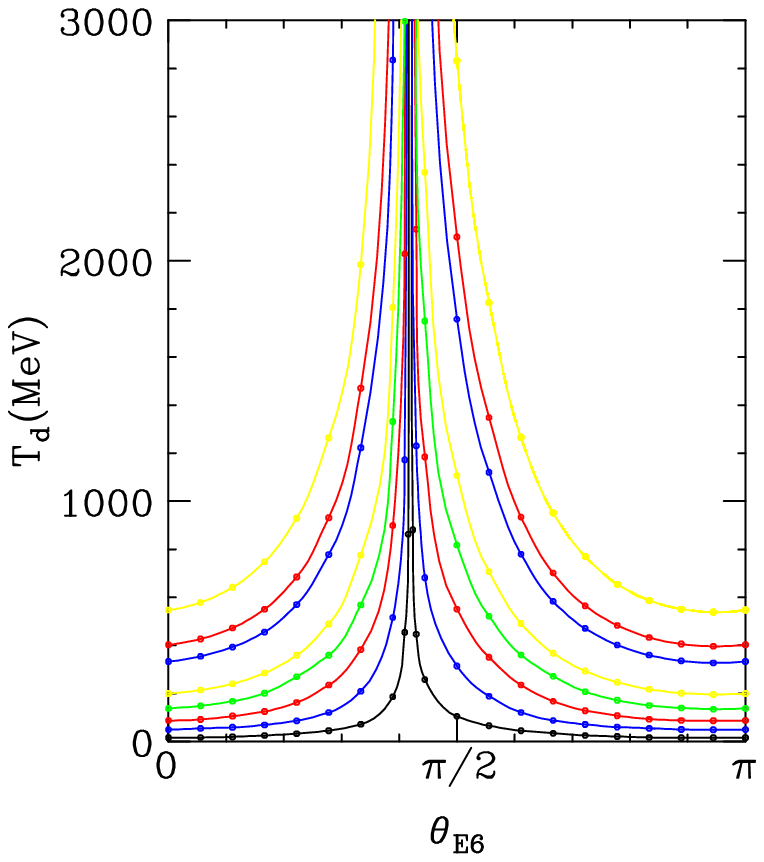}}
\resizebox{!}{2.38in}{\includegraphics{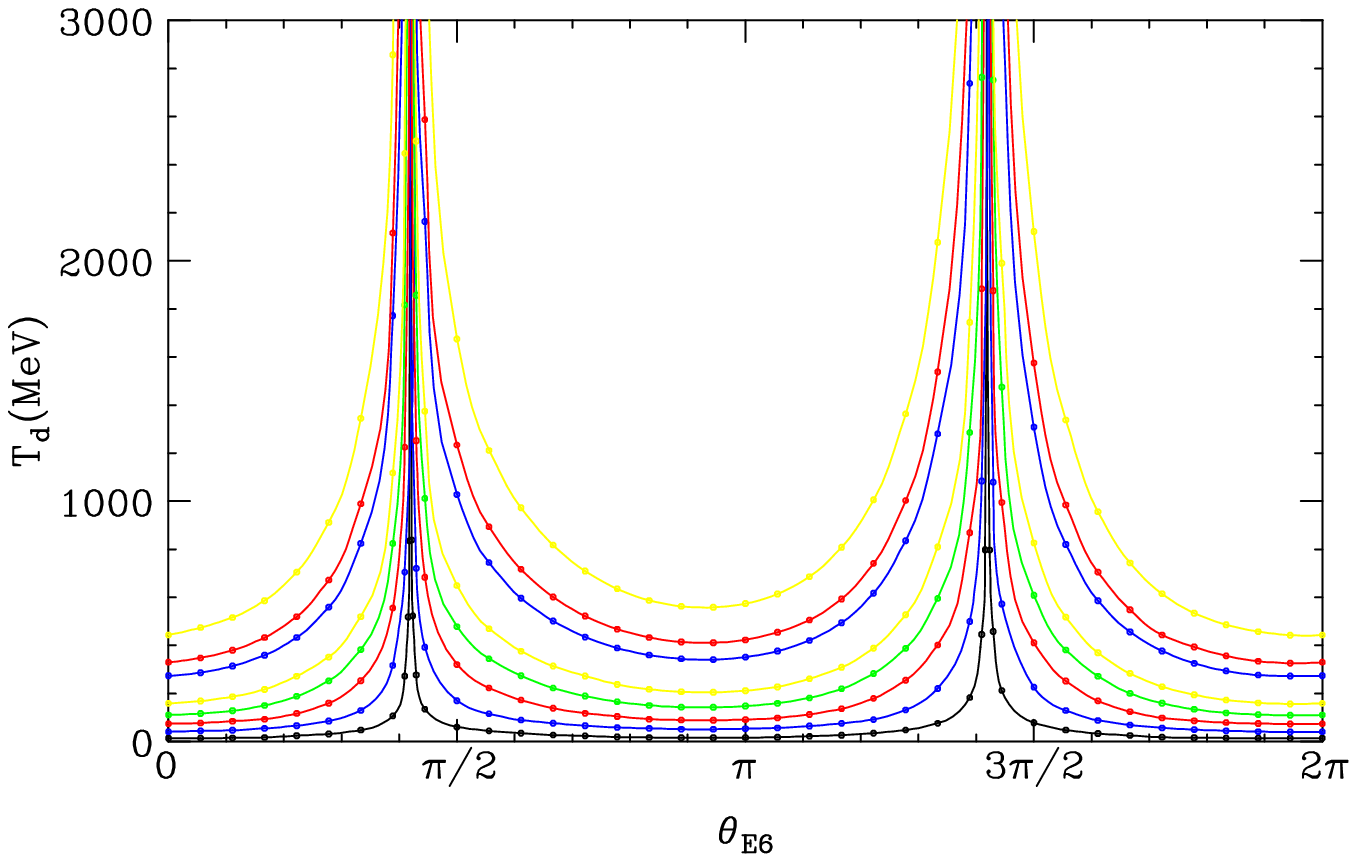}}
\resizebox{!}{2.38in}{\includegraphics{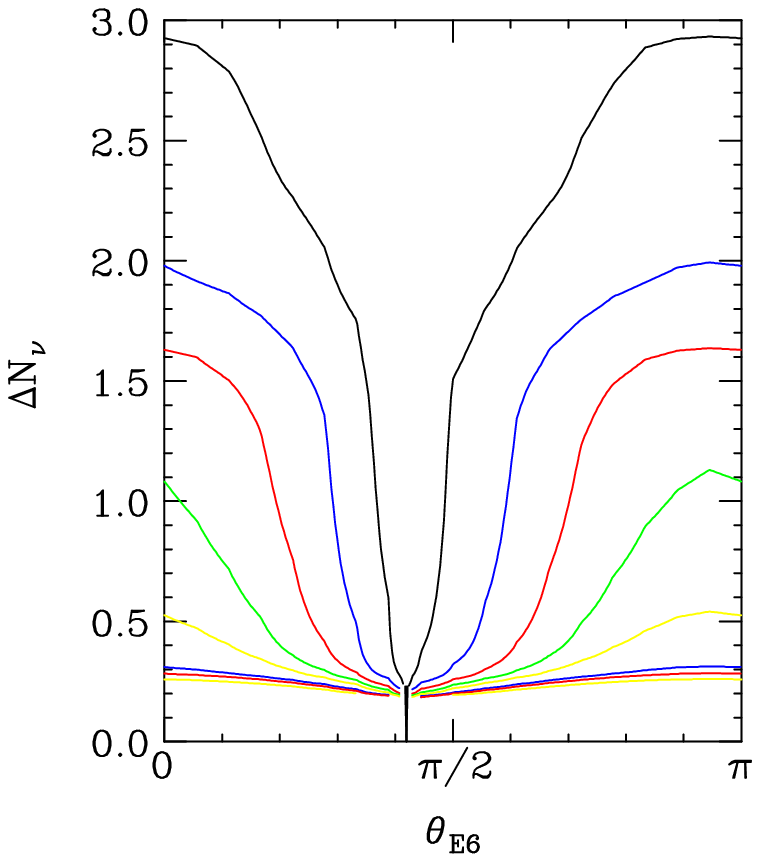}}
\resizebox{!}{2.38in}{\includegraphics{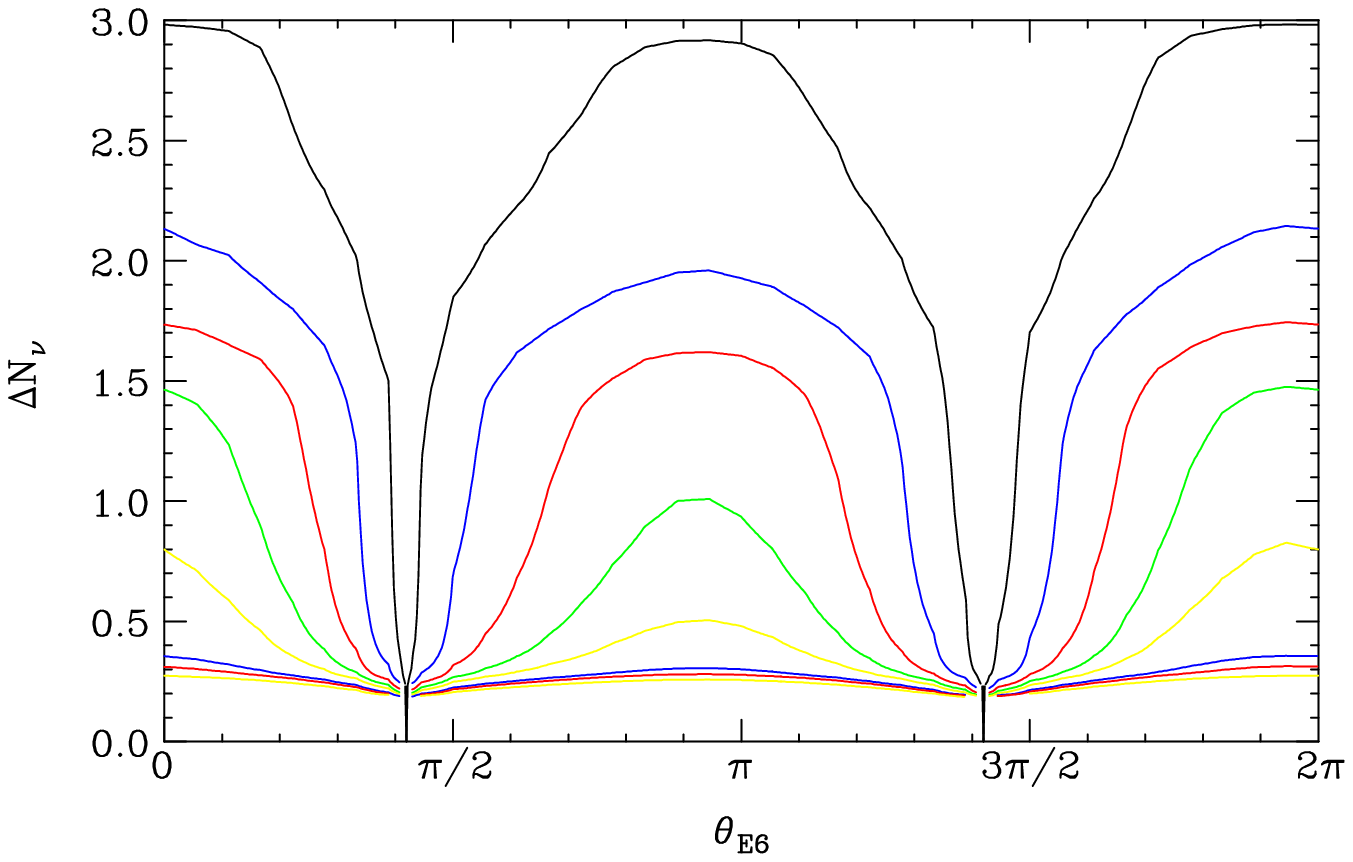}}
\resizebox{!}{2.38in}{\includegraphics{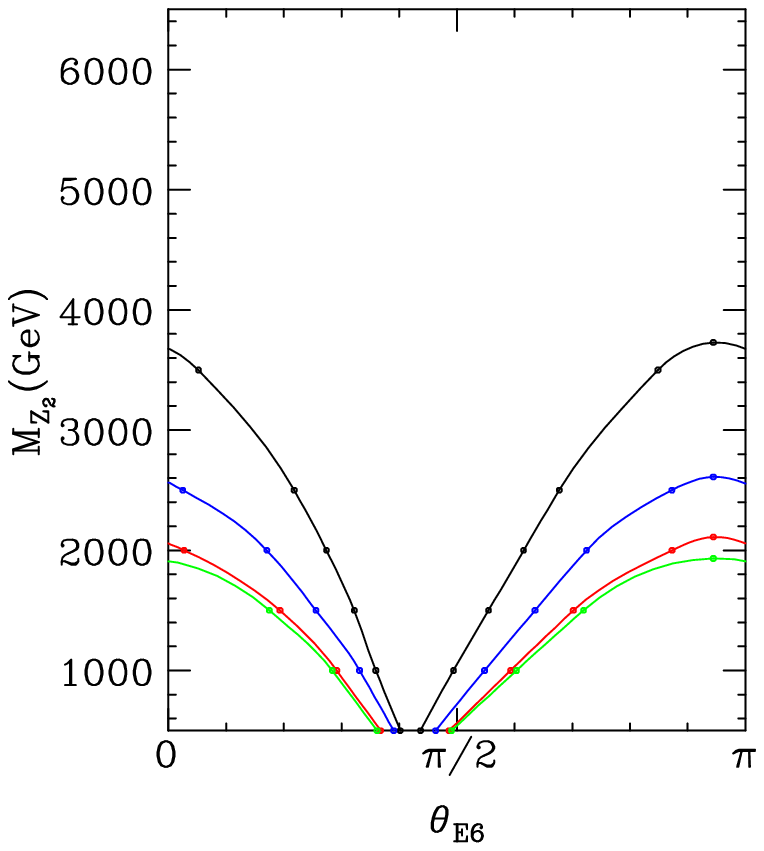}}
\resizebox{!}{2.38in}{\includegraphics{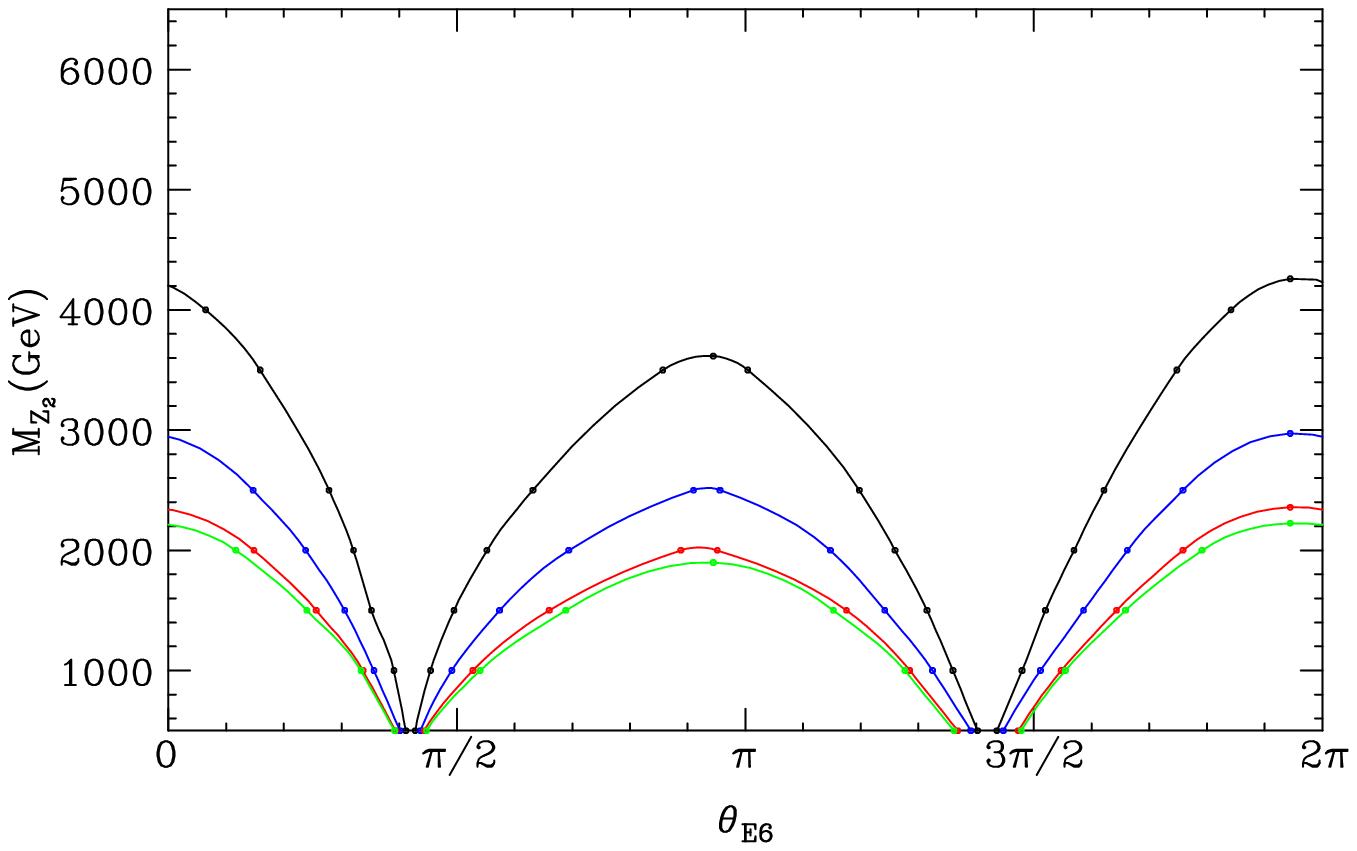}}
\end{center}
\caption{$T_d$ (top) and $\Delta N_\nu$ (middle) 
for $M_{Z_2} = $
500, 1000, 1500, 2000,  2500, 3500, 4000, and 5000 GeV, 
for $T_c = 150$ MeV and constituent masses.
Larger $M_{Z_2}$
corresponds to  higher $T_d$
and  smaller $\Delta N_\nu$.
The graphs on the left are for no mixing (case A0 in (\ref{cases})),
while the right-hand graphs are for the mass-mixing relation
$|\delta | < 0.0051/M_{Z_2}^2$ (case A1). 
The bottom graphs are $M_{Z_2}$ corresponding to
 $\Delta N_\nu = 0.3$, $0.5$, $1.0$ and $1.2$, with
 larger $\Delta N_\nu$ corresponding to smaller $M_{Z_2}$.}
\label{A1-150}
\end{figure}

\begin{figure}
\begin{center}
\resizebox{!}{2.38in}{\includegraphics{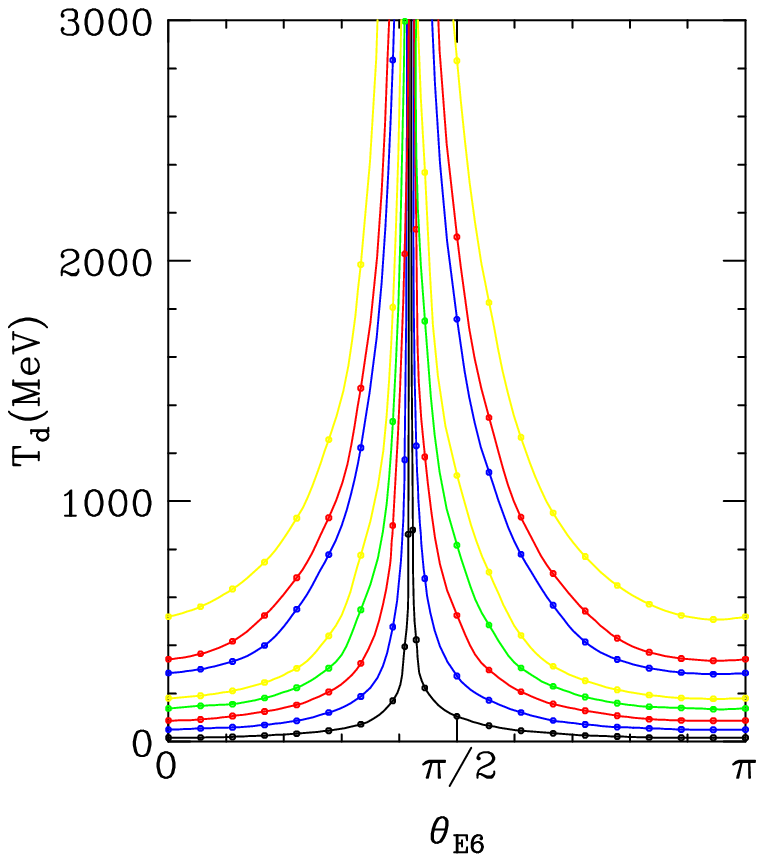}}
\resizebox{!}{2.38in}{\includegraphics{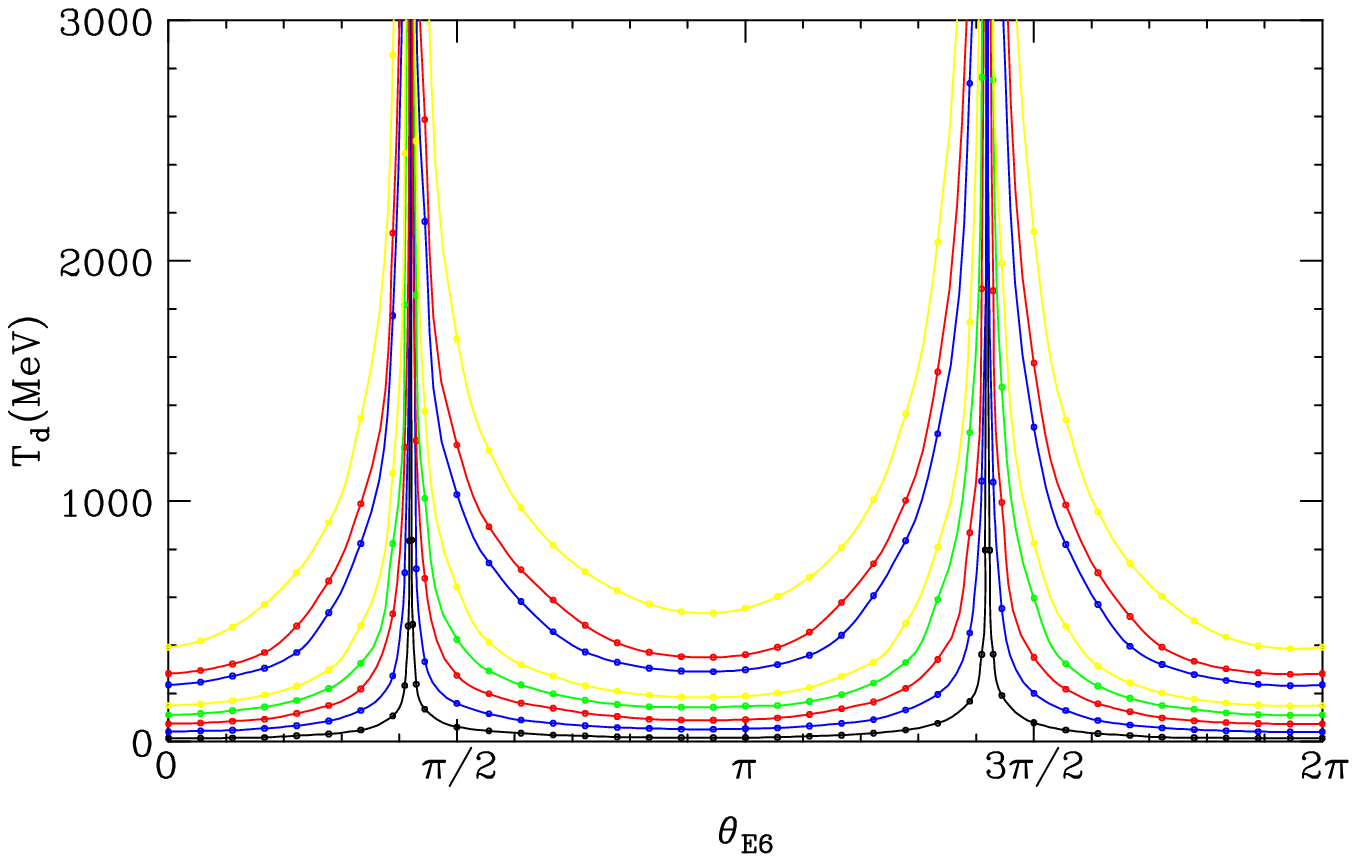}}
\resizebox{!}{2.38in}{\includegraphics{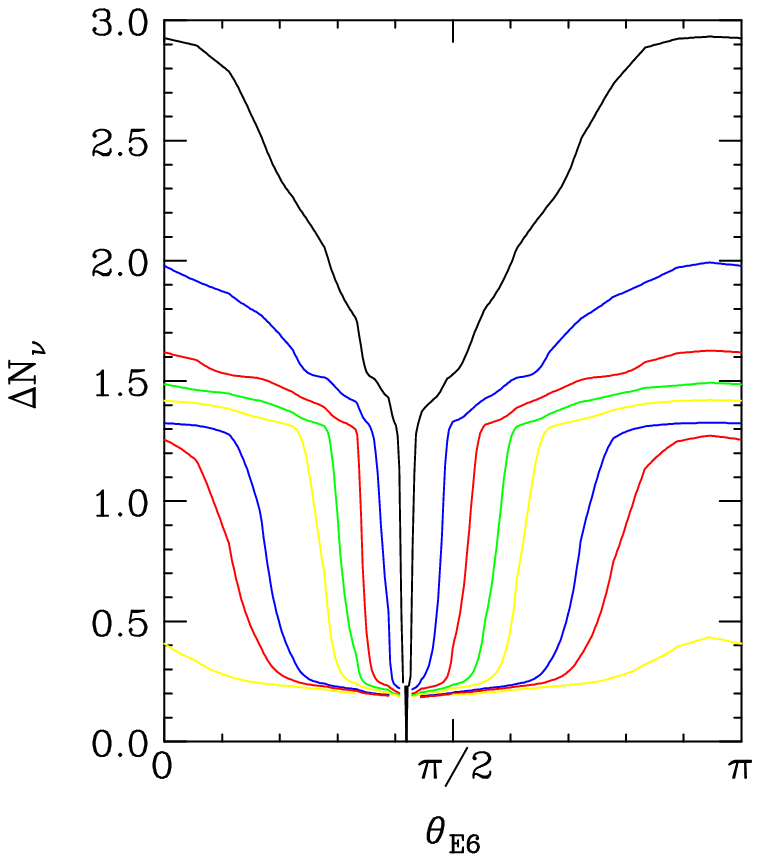}}
\resizebox{!}{2.38in}{\includegraphics{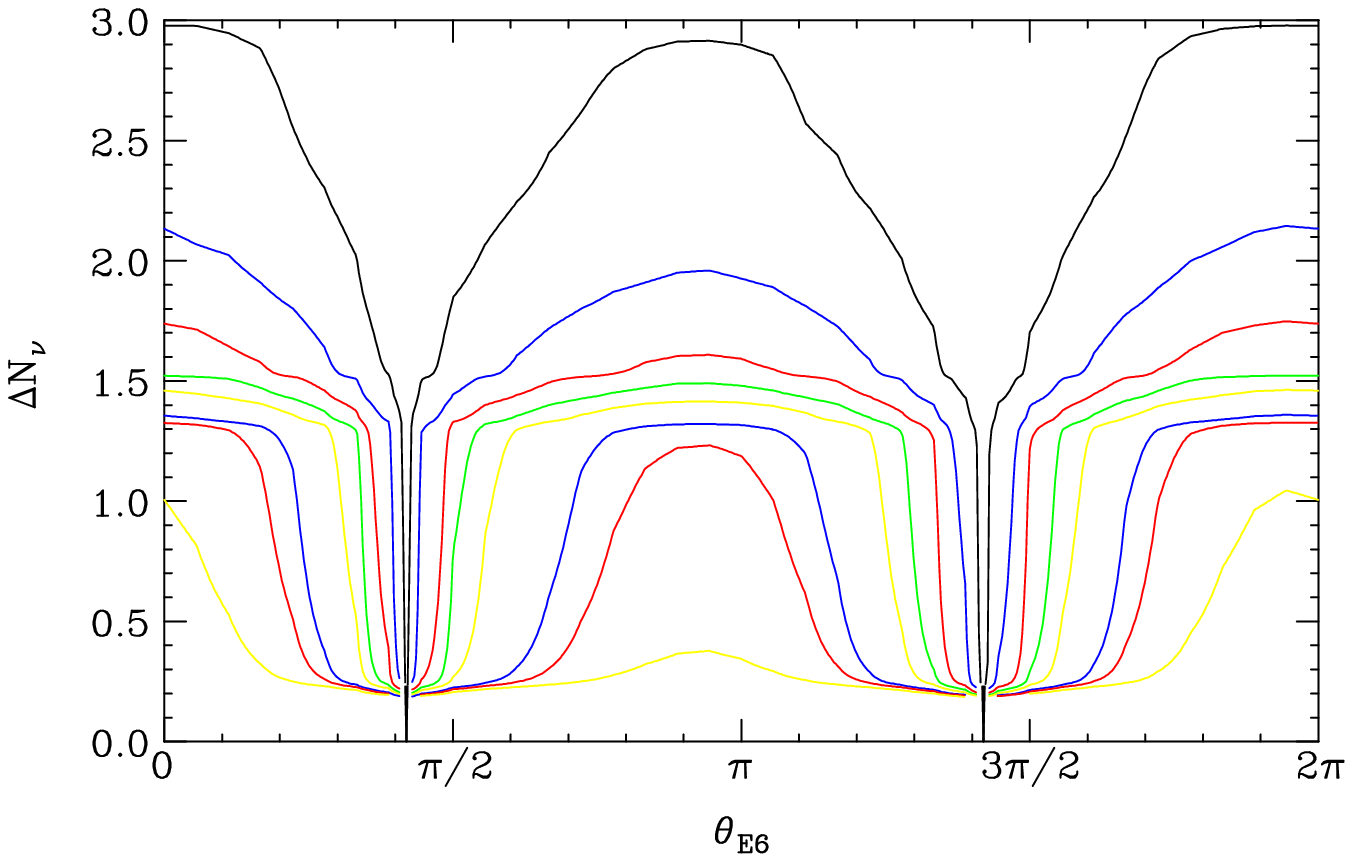}}
\resizebox{!}{2.38in}{\includegraphics{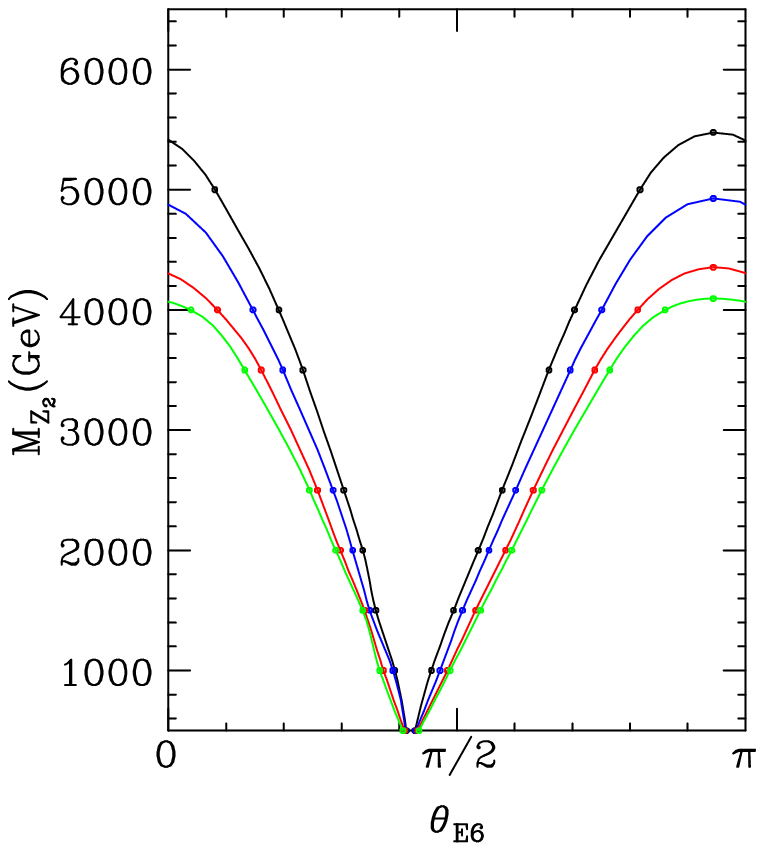}}
\resizebox{!}{2.38in}{\includegraphics{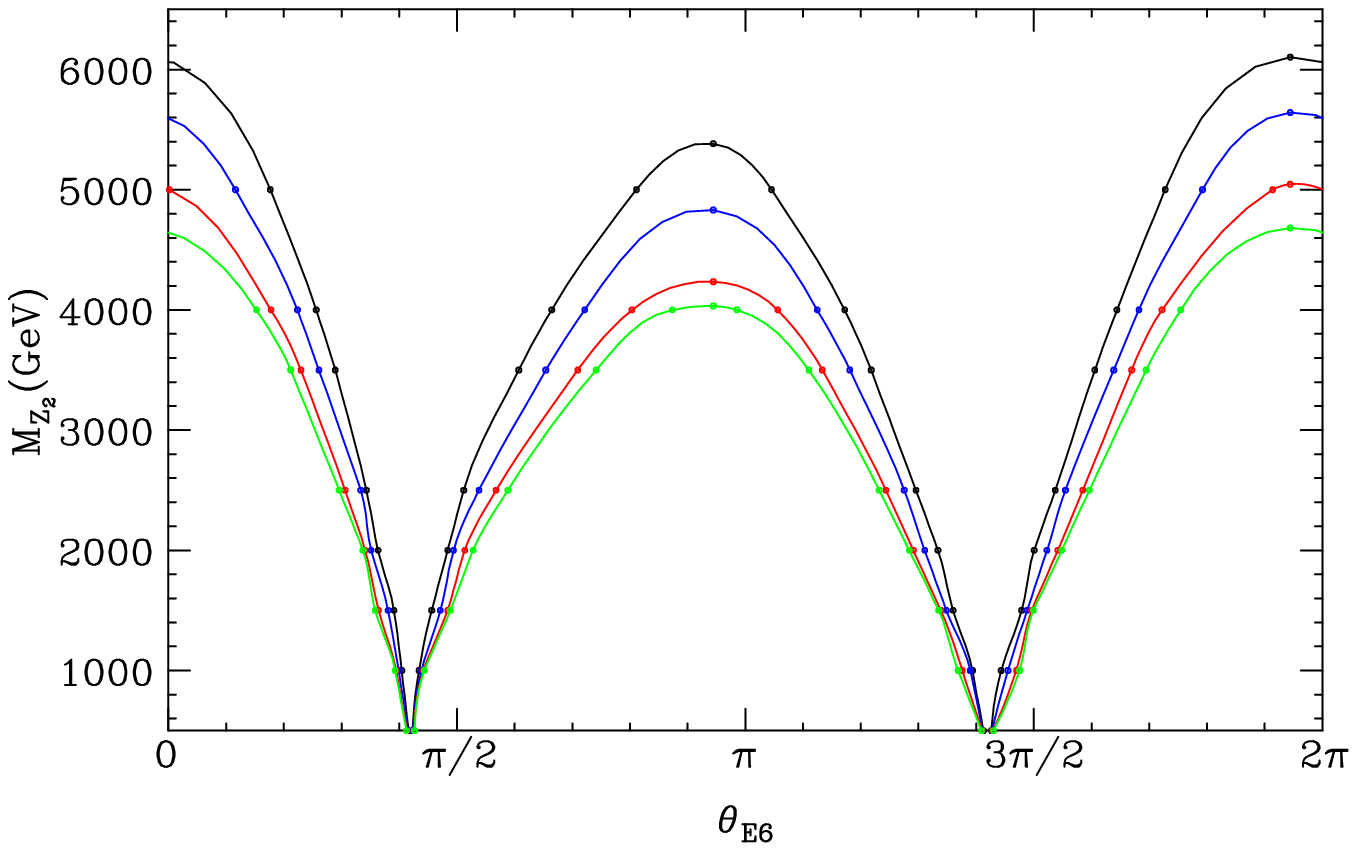}}
\end{center}
\caption{Same as Figure~\ref{A1-150}, except $T_c=400$ MeV. $T_d$ is slightly smaller
(for $T_d>$ 150 MeV) for fixed $M_{Z_2}$
and $\theta_{E6}$, while $\Delta N_\nu$ and the bound on $M_{Z_2}$ for fixed $\Delta N_\nu$
are increased.}
\label{A1-400}
\end{figure}

In this section, we present the numerical results from the calculation.
The marked points in Figures \ref{decoupling1}-\ref{A1-400}
are the results of the actual calculation, while the curves
interpolate.

Figures \ref{decoupling1} and  \ref{decoupling1a} show how the right-handed 
neutrino decoupling temperature $T_d$ and the equivalent 
number of extra neutrino species $\Delta
N_\nu$ change with $M_{Z_2}$ for  $\theta_{E6} = 2 \pi - \tan^{-1} \sqrt{\frac{5}{3}} 
\sim 1.71
\pi$ (the $\eta$ model) for constituent and current masses, respectively,
for $T_c=$ 150 and 400 MeV and the various assumptions concerning the $Z-Z'$ mixing
listed in (\ref{cases}). The no-mixing curves (A0) exhibit an approximate $T_d \sim
(M_{Z_2}/M_Z)^{4/3}$ dependence, in agreement with the simple estimate in the
Introduction~\cite{steigman,steigman2}. This is to be roughly expected because
of the $M_{Z_2}^{-4}$ dependence of the cross section for no mixing, but is not
exact because additional channels which affect both the expansion and interaction
rates open up at higher temperatures. The no-mixing curves in Figures
\ref{decoupling1} and  \ref{decoupling1a}
are reasonably described by (\ref{simple}) 
for $T_d(\nu_L) \sim$ 3 MeV for the $\eta$ model,
but the coefficients in front
of $(M_{Z_2}/M_Z)^{4/3}$ are strongly model dependent,
as is apparent in Figures \ref{A1-150}-\ref{A1-400}. $T_d$ is 
usually lower in the cases involving
$Z-Z'$ mixing, because the $Z$ annihilation channel yields a contribution 
proportional to $\delta^2$ even for infinite $M_{Z_2}$. 
That is why the (theoretically unrealistic) curves A3 for fixed $|\delta| = 0.002$
are asymptotically flat for large $M_{Z_2}$. Case A1, in which
$|\delta| \sim 0.0051/M_{Z_2}^2$, also has $T_d \sim
(M_{Z_2}/M_Z)^{4/3}$, though with a smaller coefficient than for no mixing\footnote{The
coefficient is smaller for most but not all values of $\theta_{E6}$.}, while 
A2, with $|\delta |= 0.0029/M_{Z_2}$, has $T_d \sim
(M_{Z_2}/M_Z)^{2/3}$. For case A1, $T_d$ is asymmetric under $\delta \rightarrow - \delta$
for all $M_{Z_2}$, as is apparent from (\ref{cases}) 
and (\ref{grx}). The difference vanishes
 asymptotically for A2  and A3, but even for $M_{Z_2}=$ 5 TeV there is still a difference,
especially for A2.

The decoupling temperature is
slightly lower for $T_c=$ 400 MeV than for 150 MeV, provided
it is in the range for
which the two curves in Figure~\ref{dof} differ. Both the
expansion and annihilation rates are
smaller for $T_c=$ 400 MeV, but the effect on the
expansion rate is more important
because of the gluonic degrees of freedom. Similarly,
$T_d$ is smaller for current quark masses than for constituent masses, provided
$T_d > T_c$, because of the larger annihilation 
rate\footnote{The difference between current and constituent masses
 would be reduced if
their effects in the annihilation rate were properly correlated 
with those in the expansion rate. However, as described in
Section \ref{nucleosynthesis}, the effect on $g(T)$ is small compared
with the uncertainty from $T_c$, and will be neglected.}.

The $\Delta
N_\nu$ curves change rapidly when $T_d$ reaches the quark-hadron phase 
transition temperature
$T_c$, where $g(T)$ changes significantly. That is why \dnn \ is so much larger for
$T_c$ = 400 MeV than for 150 MeV. For the no-mixing case, the difference
is significant for $M_{Z_2} \simle $ 4 TeV, and it persists to even higher masses
for the mixing cases (and to infinite mass for maximal mixing).
The only significant difference between the constituent and current quark masses
is in the maximal mixing case with $T_c=$ 150 MeV. That is because $T_d$ is very close
to $T_c$, and even a small change in $T_d$ leads to a 
significant change in $g(T)$, as can be seen in Figure~\ref{dof}.

It is apparent from Figures \ref{decoupling1} and  \ref{decoupling1a}
that the $\eta$ model leads to a significant \dnn \ for all of the
cases and parameter ranges considered. Even the very conservative constraint
$\Delta N_\nu < 1$ implies $M_{Z_2} > 1.5-2.2$ TeV for $T_c = 150$ MeV,
or, limiting ourselves to the most realistic cases A0 and A1, $M_{Z_2} > 1.5-1.9$ TeV.
For $T_c = 400$ MeV one finds
 $M_{Z_2} > 3.3-4$ TeV for A0 and A1, $M_{Z_2} > 5$ TeV for A2 
and no allowed values for A3.
All of these are much more stringent than the direct laboratory limit 
of 620 GeV~\cite{explim} or the indirect limits from precision electroweak
data~\cite{indirect}. The more stringent limit 
$\Delta N_\nu < 0.3$ is satisfied for cases A0 and A1 
for  $M_{Z_2} > 2.5-3.2$ TeV for $T_c = 150$ MeV,
and $M_{Z_2} > 4.0-4.9$ TeV for $T_c = 400$ MeV. It is not satisfied for 
case A2 with $T_c=$ 400 MeV until extremely high masses, and  never for (fixed)
maximal mixing unless one takes a mixing much smaller than the 
 present accelerator limit ($|\delta| < 0.0024$) \cite{LEPmixing}.

Figures \ref{A1-150} and \ref{A1-400} display the results for the class of $E_6$ models
parametrized by the angle $\theta_{E6}$ defined in (\ref{Qdef}), for
constituent masses and
$T_c=$ 150 MeV and 400 MeV, respectively. Each  figure
includes the no-mixing case and the mixing assumption A1 defined in (\ref{cases}),
which is the most stringent and realistic.
The limits in the presence of $Z-Z'$ mixing are asymmetric under
$\delta \rightarrow - \delta$. This is represented in the right-handed graphs
by taking $\delta < 0$ but allowing $\theta_{E6}$ to run from 0 to $2\pi$,
so that the $(\pi-2\pi)$ range for $\delta < 0$ is equivalent to 
$(0-\pi)$ with $\delta > 0$.
The top graphs display $T_d$ as a function of $\theta_{E6}$
for fixed values $M_{Z_2}=$ 500, 1000, 1500, 2000,  2500, 3500, 4000,
and 5000 GeV, with  larger
$M_{Z_2}$ corresponding to  higher $T_d$.
The middle graphs show \dnn \ as a function of $\theta_{E6}$ for the same
values of $M_{Z_2}$, with larger $M_{Z_2}$
corresponding to smaller \dnn. The bottom figures show the lower bounds on $M_{Z_2}$ for
$\Delta N_\nu < 0.3$, $0.5$, $1.0$ and $1.2$, with
 larger $\Delta N_\nu$ corresponding to smaller $M_{Z_2}$.

It is seen that $T_d$ becomes very large and the $M_{Z_2}$ limits essentially disappear 
as $\theta_{E6}$ approaches  $\theta_{E6} \sim 0.42
\pi$ or $ 1.42 \pi$, for which $\nu_R$ decouples completely 
($Q(\nu_R) = 0$), but the details depend on
the new physics at the electroweak and higher scales 
(we only explicitly included particles up to the
$b$ quark).
$\theta_{E6} = 1.71 \pi$ corresponds to the $\eta$ model with $\delta < 0$,
while $\theta_{E6} = 0.71 \pi$ corresonds to $\delta > 0$. It is seen from the figures
that \dnn \ is larger for values of $\theta_{E6}$ closer to 0 (the $\chi$ model),
but are weaker near $\theta_{E6}=\pi/2$ (the $\psi$ model).

From the figures it is apparent that  requiring $\dnn \le 1$ 
excludes much of the interesting
parameter space for $T_c=$ 150 MeV, except for large $Z_2$ masses or regions
 very close to the $\nu_R$ decoupling angles 
$\sim 0.42\pi$ and $1.42 \pi$. In particular, the $\dnn \le 1$ constraint is satisfied
for all values of $\theta_{E6}$   for $M_{Z_2} \simgr 2.2$ TeV if there is no
mixing, with a slightly more stringent constraint $M_{Z_2} \simgr 2.4$ TeV for mixing
assumption A1. The corresponding $M_{Z_2}$ limits
  for $\dnn \le 0.3$ are 3.8 and 4.3 TeV.
The constraints for $T_c=$ 400 MeV are even more stringent, essentially requiring
$\nu_R$ decoupling or very  large $Z_2$ masses. One has
 $\dnn \le 1 (0.3)$ for all $\theta_{E6}$ for cases A0 and A1 for
$M_{Z_2} \simgr 5.1(6.1)$ TeV.

\section{Discussion and Conclusion} 
\label{discussion}

Many theories beyond the standard model predict the existence of additional
\zpr \ gauge bosons at the TeV scale. The associated \upr \ gauge symmetry
often prevents the large Majorana masses needed for an ordinary neutrino seesaw
model.  One possibility is that the neutrino masses are Dirac and small.
In that case, there is a possibility of producing the  sterile
``right-handed'' neutrino partners $\nu_R$  via \zpr \ interactions prior to 
nucleosynthesis~\cite{steigman,steigman2}, leading to a faster expansion and 
additional $^4He$.

We have studied the  right-handed neutrino decoupling temperature $T_d$ 
in a class of $E_6$-motivated \upr \ models as a function of the \zpr \ mass
and couplings (determined by an angle $\theta_{E6}$) for a variety of assumptions
concerning the $Z-Z'$ mixing angle $\delta$, the quark-hadron transition temperature $T_c$, and the
nature (constituent or current) of the quark masses. We have taken all relevant
 channels (quark, gluon, lepton, and hadron) into account, not only in
the expansion rate $H(T)$ and entropy, but also in the  rate
$\Gamma(T)$ for a massless right-handed neutrino
pair to annihilate into a fermion or pion pair via the ordinary or heavy $Z$ bosons.
We therefore obtain a larger annihilation rate, and thus a lower decoupling
temperature and more stringent constraints, than earlier calculations, which
only included annihilation into $e^+e^-$ and   $\nu_L \overline{\nu_L}$.

From the decoupling temperature and  entropy conservation as quarks and gluons are
confined or as various heavy particle
types decouple and annihilate, one can obtain the  number of right-handed neutrinos
at nucleosynthesis, expressed in terms of the equivalent number  $\Delta N_\nu$ of new
ordinary neutrino species,
for various sets
of model parameters $M_{Z_2}$, $\delta$, $\theta_{E6}$, and $T_c$.
Most recent studies of the primordial abundances obtain upper limits on $\Delta N_\nu$ 
in the range (0.3--1)~\cite{bbnreview,lisi}. As can be seen in  Figures
\ref{A1-150}-\ref{A1-400}, this implies rather
stringent constraints on the \zpr \ parameters for most values of $\theta_{E6}$.
For $T_c=$ 150 MeV, the constraint $\dnn < 0.3 (1)$ is satisfied for
all $\theta_{E6}$ for $M_{Z_2} \simgr 3.8 (2.2)$ TeV for no $Z-Z'$ mixing,
and for $M_{Z_2} \simgr 4.3 (2.4)$ TeV allowing the range of mixing angles $\delta$
obtained approximately when one assumes that the scalar fields responsible for the
mixing are contained in the 27 or $\overline{27}$-plet of $E_6$ (case A1 in (\ref{cases})).
For $T_c=$ 400 MeV the  constraints are much stronger,
$M_{Z_2} \simgr 6.1 (5.1)$ TeV for $\dnn < 0.3 (1)$. The strong dependence on
$T_c$ is due to the large increase in the number of degrees of freedom
for temperatures $\simgr T_c$ (Figure (\ref{dof})), so that the number density
of $\nu_R$ is strongly diluted for $T_d \simgr T_c$.
The constraints are strongest for $\theta_{E6}$ close to 0 or $\pi$,
i.e., near the $\chi$ model, which corresponds to 
$ SO(10)  \to SU(5) \times U(1)_\chi$, and are very weak near the
$\psi$ model corresponding to $E_6 \to SO(10) \times U(1)_\psi$,  $\theta_{E6}=\pi/2$.
They disappear entirely at the values 
$\theta_{E6} = 0.42\pi$ and $1.42 \pi$, for which the $\nu_R$
decouple from the \zpr. The often considered $\eta$ model,
$\theta_{E6} = 2 \pi - \tan^{-1} \sqrt{\frac{5}{3}}=1.71 \pi$ (or $0.71 \pi$ for $-Z_\eta$)
is somewhere in between, with the constraints shown in more detail in
Figures \ref{decoupling1} and \ref{decoupling1a}.

Except near the $\nu_R$ decoupling angles, the \zpr \ mass and mixing constraints
from nucleosynthesis are much more stringent than the existing laboratory
limits from searches for direct production or from precision electroweak
data, and are comparable to the ranges that may ultimately be probed
at proposed colliders.
They are qualitatively similar to
the limits from energy emission from Supernova 1987A~\cite{supernova},
but somewhat more stringent for
$\dnn<0.3$, and have entirely different theoretical and systematic uncertainties.

There are several ways to evade the nucleosynthesis constraints on an extra \zpr.
One possibility is to generate small Majorana neutrino masses for the ordinary
neutrinos by invoking an extended seesaw model~\cite{TEVseesaw}, in which the extra
sterile neutrinos are typically at the TeV scale. Another possibility
is that the $\nu_R$ decouple from the \zpr, in which case the constraints
disappear. This can in fact occur naturally in classes of models in which
one combination of  the $\chi$ and $\psi$ charges is broken at a
large scale associated with an $F$ and $D$-flat direction~\cite{decouplingmodel},
leaving a light \zpr \ which decouples from the $\nu_R$\footnote{A large Majorana
mass for the $\nu_R$ may still be forbidden in the model.}. Yet
another possibility is to weaken the observational constraint
on \dnn \ by allowing a large  excess\footnote{Several authors~\cite{equilibrate} have 
recently argued
that the observed atmospheric and solar neutrino mixing (for the 
large mixing angle solution)
would equilibrate the $\nu_e, \ \nu_\mu$, and $\nu_\tau$ asymmetries. 
They obtain stringent
constraints on the asymmetry if one requires a balance between
 the effects on the $\nu_e n
\leftrightarrow e^- p$ and expansion rates. However, these limits do not apply 
in the present case because of the
additional contribution to the expansion rate from the $\nu_R$.}
 of $\nu_e$ with respect to $\bar{\nu}_e $. This would, however, require a somewhat
fine-tuned cancellation between the effects of the $\nu_R$ and 
the $\nu_e - \bar{\nu}_e$ asymmetry.

Similar constraints
on the $W'$ and $Z'$ properties in $SU(2)_L \times SU(2)_R \times U(1)$ models~\cite{LR}
are under investigation~\cite{lrconstraints}.

\section*{Acknowledgments}
This research was supported in part by the U.S. Department of 
Energy under Grants No. EY-76-02-3071 and No. DE-FG02-95ER40896, and 
in part by the Wisconsin
Alumni Research Foundation. 
VB thanks the Kavli Institute for Theoretical
Physics at the University of California in Santa Barbara for hospitality.

\newpage

\newpage

\begin{thebibliography}{99}

\bibitem{ell}
For  recent surveys, see
J.~Erler, P.~Langacker and T.~J.~Li,
Phys.\ Rev.\ D {\bf 66}, 015002 (2002);
S.~Hesselbach, F.~Franke and H.~Fraas,
Eur.\ Phys.\ J.\ {\bf C23}, 149 (2002).


\bibitem{string}
M.~Cveti\v c and P.~Langacker,
Phys.\ Rev.\ D {\bf 54}, 3570 (1996) and
Mod.\ Phys.\ Lett.\ A {\bf 11}, 1247 (1996).

\bibitem{review}
For a review, see, M.~Cveti\v c and P.~Langacker,
in {\it Perspectives on supersymmetry}, ed. G. L. Kane
(World, Singapore, 1998), p. 312.

\bibitem{DSB}
For a review, see
C.~T.~Hill and E.~H.~Simmons,
hep-ph/0203079.

\bibitem{muprob}
D.~Suematsu and Y.~Yamagishi,
Int.\ J.\ Mod.\ Phys.\ A {\bf 10}, 4521 (1995);
M.~Cveti\v c, D.~A.~Demir, J.~R.~Espinosa, L.~L.~Everett and P.~Langacker,
Phys.\ Rev.\ D {\bf 56}, 2861 (1997)
[Erratum-ibid.\ D {\bf 58}, 119905 (1997)].

\bibitem{general}
J.~Erler,
Nucl.\ Phys.\ B {\bf 586}, 73 (2000).

\bibitem{explim}  F. Abe { et al.} [CDF Collaboration],
{ Phys. Rev.  Lett.} {\bf 79}, 2192 (1997).

\bibitem{LEPmixing}
R.~Barate {\it et al.}  [ALEPH Collaboration],
Eur.\ Phys.\ J.\ C {\bf 12}, 183 (2000);
P.~Abreu {\it et al.}  [DELPHI Collaboration],
Phys.\ Lett.\ B {\bf 485}, 45 (2000).

\bibitem{indirect}
J.~Erler and P.~Langacker,
Phys.\ Lett.\ B {\bf 456}, 68 (1999), and references therein.

\bibitem{APV} C.S. Wood { et al.}, { Science} {\bf 275}, 1759
(1997);
S.C. Bennett and C.E. Wieman, { Phys. Rev. Lett.} {\bf 82}, 2484
(1999). 

\bibitem{interp}
V.A. Dzuba, V.V. Flambaum, and J.S.M. Ginges, hep-ph/0204134;
M. Y. Kuchiev,
J.\ Phys.\ BB {\bf 35}, L503 (2002).


\bibitem{NuTeV}
G.~P.~Zeller { et al.}  [NuTeV Collaboration],
Phys.\ Rev.\ Lett.\  {\bf 88}, 091802 (2002).



\bibitem{hints}
R.~Casalbuoni, S.~De Curtis, D.~Dominici and R.~Gatto,
Phys.\ Lett.\ B {\bf 460}, 135 (1999);
J.~L.~Rosner,
Phys.\ Rev.\ D {\bf 61}, 016006 (2000);
J.~Erler and P.~Langacker,
Phys.\ Rev.\ Lett.\  {\bf 84}, 212 (2000).

\bibitem{collider}
For reviews, see
M.~Cveti\v c and S.~Godfrey,
hep-ph/9504216;
A.~Leike,
Phys.\ Rept.\  {\bf 317}, 143 (1999).
For a recent update, see S.~Godfrey
in {\it Proc. of the APS/DPF/DPB Summer Study
on the Future of Particle Physics (Snowmass 2001) } ed. N.~Graf,
hep-ph/0201093 and hep-ph/0201092.

\bibitem{seesaw}
M. Gell-Mann, P. Ramond, and R. Slansky, in {\it
Supergravity}, ed. F. van Nieuwenhuizen and D. Freedman, (North
Holland, Amsterdam, 1979) p. 315; T. Yanagida, {\it Proc. of the
Workshop on Unified Theory and the Baryon Number of the Universe},
KEK, Japan, 1979; S. Weinberg, 
Phys.\ Rev.\ Lett.\  {\bf 43}, 1566 (1979);
R.~N.~Mohapatra and G.~Senjanovic,
Phys.\ Rev.\ Lett.\  {\bf 44}, 912 (1980) and
Phys.\ Rev.\ D {\bf 23}, 165 (1981).

\bibitem{TEVseesaw}
See, for example,
R.~N.~Mohapatra and J.~W.~Valle,
Phys.\ Rev.\ D {\bf 34}, 1642 (1986), and references therein;
for an extension to \upr \ models,
see
A.~E.~Faraggi,
Phys.\ Lett.\ B {\bf 245}, 435 (1990);
J. Kang, P. Langacker, and T. Li, UPR-1010-T, to appear.

\bibitem{HDO}
These could arise, for example, in a variant on the
model in 
P.~Langacker,
Phys.\ Rev.\ D {\bf 58}, 093017 (1998).

\bibitem{LED}
See, for example,
K.~R.~Dienes, E.~Dudas and T.~Gherghetta,
Nucl.\ Phys.\ B {\bf 557}, 25 (1999);
G.~R.~Dvali and A.~Y.~Smirnov,
Nucl.\ Phys.\ B {\bf 563}, 63 (1999);
N. Arkani-Hamed, S. Dimopoulos, G. R. Dvali and J. March-Russel,
Phys.\ Rev.\ {\bf D65}, 024032 (2002);
T. Appelquist, B. Dobrescu, E. Ponton and H. Yee,
Phys.\ Rev.\ {\bf D65}, 105019 (2002);
H.~Davoudiasl, P.~Langacker and M.~Perelstein,
Phys.\ Rev.\ D {\bf 65}, 105015 (2002),
and references therein.
\bibitem{LR}
J. C. Pati and A. Salam,
Phys.\ Rev.\ {\bf D10}, 275 (1974);
R. N. Mohapatra and J. C. Pati,
Phys.\ Rev.\ {\bf D11}, 566 (1975);
ibid. Phys.\ Rev.\ {\bf D11}, 2558 (1975);
G. Senjanovic and R. N. Mohapatra,
Phys.\ Rev.\ {\bf D12}, 1502 (1975);
R.~N.~Mohapatra,
{\it Unification And Supersymmetry} (Springer, New York, 1986).

\bibitem{steigman}
G.~Steigman, K.~A.~Olive and D.~N.~Schramm,
Phys.\ Rev.\ Lett. {\bf 43}, 239 (1979).

\bibitem{steigman2}
K.~A.~Olive and D.~N.~Schramm and G.~Steigman,
Nucl.\ Phys.\ {\bf B180}, 497 (1981).

\bibitem{dolgov}
For a general review of neutrinos in cosmology, see
A.~D.~Dolgov,
Phys.\ Rept.\ {\bf 370}, 333 (2002) 



\bibitem{supernova}
G. Raffelt and D. Seckel,
Phys.\ Rev.\ Lett. {\bf 60}, 1793 (1988);
R. Barbieri and R. N. Mohapatra,
Phys.\ Rev.\ {\bf D39}, 1229 (1989);
J. A. Grifols and E. Masso,
Nucl.\ Phys.\ {\bf B331}, 244 (1990);
J.~A.~Grifols, E.~Masso and T.~G.~Rizzo,
Phys.\ Rev.\ D {\bf 42}, 3293 (1990);
T.~G.~Rizzo,
Phys.\ Rev.\ D {\bf 44}, 202 (1991).

\bibitem{yang}
J.~Yang, D.~N.~Schramm, G.~Steigman and R.~T.~Rood,
Astrophys.\ J.\ {\bf 227}, 697 (1979).

\bibitem{bbnreview}
For  recent reviews, see G. Steigman, astro-ph/0009506;
 the  articles by K. A. Olive and J. A. Peacock and by 
B. D. Fields and S. Sarkar
in~\cite{PDG}; and \cite{dolgov}.

\bibitem{lisi}
E.~Lisi, S.~Sarkar and F.~L.~Villante,
Phys.\ Rev.\ D {\bf 59}, 123520 (1999).

\bibitem{PDG}
Particle Data Group: K. Hagiwara et al, Phys.\ Rev.\ {\bf D66}, 010001 (2002). 

\bibitem{prev1}
J.~R.~Ellis, K.~Enqvist, D.~V.~Nanopoulos and S.~Sarkar,
Phys.\ Lett.\ B {\bf 167}, 457 (1986);
J.~L.~Lopez and D.~V.~Nanopoulos,
Phys.\ Lett.\ B {\bf 241}, 392 (1990).

\bibitem{prev2}
M.~C.~Gonzalez-Garcia and J.~W.~Valle,
Phys.\ Lett.\ B {\bf 240}, 163 (1990).

\bibitem{afdvn}
A.~E.~Faraggi and D.~V.~Nanopoulos,
Mod.\ Phys.\ Lett.\ A {\bf 6}, 61 (1991).

\bibitem{ETC}
L.~M.~Krauss, J.~Terning and T.~Appelquist,
Phys.\ Rev.\ Lett.\  {\bf 71}, 823 (1993).

\bibitem{lw}
For a study of \upr \ breaking in supersymmetric $E_6$ models, 
see
P.~Langacker and J.~Wang,
Phys.\ Rev.\ {\bf D58}, 115010 (1998).

\bibitem{famnon}
P.~Langacker and M.~Plumacher,
Phys.\ Rev.\ D {\bf 62}, 013006 (2000).

\bibitem{kinetic}
B. Holdom,
Phys.\ Lett.\ {\bf B259}, 329 (1991);
F.~Del Aguila, M.~Masip and M.~Perez-Victoria,
Acta Phys.\ Polon.\ B {\bf 27}, 1469 (1996);
  K.~S.~Babu, C.~F.~Kolda and J.~March-Russell,
  Phys.\ Rev.\ D {\bf 57}, 6788 (1998).

\bibitem{stringmodels}
See, for example,
G.~Cleaver, M.~Cvetic, J.~R.~Espinosa, L.~L.~Everett, P.~Langacker and J.~Wang,
Phys.\ Rev.\ D {\bf 59}, 055005 (1999);
M.~Cvetic, P.~Langacker and G.~Shiu,
Phys.\ Rev.\ D {\bf 66}, 066004 (2002).


\bibitem{luo}
P.~Langacker and M.~X.~Luo,
Phys.\ Rev.\ D {\bf 45}, 278 (1992), and references therein.

\bibitem{erler2}
J. Erler and P. Langacker, in \cite{PDG}.

\bibitem{olive}
R.~V.~Wagoner and G.~Steigman,
Phys.\ Rev.\ D {\bf 20}, 825 (1979);
K.~A.~Olive,
Nucl.\ Phys.\ B {\bf 190}, 483 (1981) and
{\it Neutrino 79}, ed. A. Haatuft and C. Jarlskog (Astvedt Industrier
A/S, Norway, 1979), v.2:421.


\bibitem{srednicki}
M.~Srednicki, R.~Watkins and K.~A.~Olive,
Nucl.\ Phys.\ B {\bf 310}, 693 (1988). For a recent discussion, see
K. A. Olive and J. A. Peacock, \cite{PDG}.
For a  larger temperature range, see P. Gondolo and G. Gelmini,
Nucl.\ Phys.\ {\bf B360}, 145 (1991).

\bibitem{sakurai}
See,  for example J.~J.~Sakurai, {\it Currents and Mesons}, 
University of Chicago Press, 1969.

\bibitem{sakurai2}
G. J. Gounaris and J. J. Sakurai, Phys. Rev. Lett. {\bf 21}, 244 (1968);
C.~Gale and J.~Kaputsta,
Phys.\ Rev.\ {\bf C35}, 2107 (1987).

\bibitem{Kolb}
See, for example, 
E.~W.~Kolb and M.~S.~Turner,
{\it The Early Universe} (Addison-Wesley, Redwood City, 1990) and \cite{fornengo}.

\bibitem{fornengo}
N.~Fornengo, C.~W.~Kim and J.~Song,
Phys.\ Rev.\ {\bf D56}, 5123 (1997).

\bibitem{decouplingmodel}
P. Langacker, in preparation.

\bibitem{equilibrate}
C.~Lunardini and A.~Y.~Smirnov,
Phys.\ Rev.\ D {\bf 64}, 073006 (2001);
A.~D.~Dolgov, S.~H.~Hansen, S.~Pastor, S.~T.~Petcov, G.~G.~Raffelt and D.~V.~Semikoz,
Nucl.\ Phys.\ B {\bf 632}, 363 (2002);
Y.~Y.~Wong,
Phys.\ Rev.\ D {\bf 66}, 025015 (2002);
K.~N.~Abazajian, J.~F.~Beacom and N.~F.~Bell,
Phys.\ Rev.\ D {\bf 66}, 013008 (2002).

\bibitem{lrconstraints}
V. Barger, P. Langacker, and H.-S. Lee,in preparation.

\end{thebibliography}
\end{document}